\documentclass[12pt]{article}

\ifx\pdfoutput\undefined
\usepackage[dvips,bookmarks]{hyperref}
\else
\usepackage{hyperref}
\fi
\hypersetup{colorlinks=false,bookmarksopen,bookmarksnumbered,citecolor=blue,
   pdfstartview=FitH}

\usepackage{latexsym}
\usepackage{amssymb,amsfonts,amsmath}
\usepackage{graphicx} 
\usepackage{indentfirst}
\usepackage{bbm}
\usepackage{amssymb}
\usepackage{verbatim}
\usepackage{amsmath, amsthm,amssymb}
\usepackage{mathrsfs}
\usepackage{hyperref}
\usepackage{amsfonts}
\usepackage{dsfont}
\parskip=\medskipamount

\newcommand {\cC}{{\cal C}}
\newcommand {\cD}{{\cal D}}

\newcommand {\cG}{{\cal G}}
\newcommand {\cH}{{\cal H}}

\newcommand {\cJ}{{\cal J}}
\newcommand {\cK}{{\cal K}}
\newcommand {\cL}{{\cal L}}
\newcommand {\cM}{{\cal M}}
\newcommand {\cN}{{\cal N}}

\newcommand {\cR}{{\cal R}}

\def\a{\alpha}
\def\b{\beta}

\def\d{\delta}
\def\e{\epsilon}

\def\g{\gamma}
\def\G{\Gamma}

\def\l{\lambda}
\def\m{\mu}

\def\o{\omega}

\def\q{\theta}
\def\r{\rho}
\def\s{\sigma}
\def\t{\tau}

\def\x{\xi}
\def\z{\zeta}
\def\D{\Delta}

\def\L{\Lambda}
\def\O{\Omega}

\def\Q{\Theta}
\def\S{\Sigma}

\def\X{\Xi}
\def\ri{{\rm i}}

\newcommand{\ve}{\varepsilon}                            %new
\newcommand{\pa}{\partial}                           %new
\newcommand{\hf}{\frac12}
\newcommand{\be}{\begin{equation}}
\newcommand{\ee}{\end{equation}}
\newcommand{\bea}{\begin{eqnarray}}
\newcommand{\eea}{\end{eqnarray}}
\newcommand{\non}{\nonumber}
\newcommand{\ba}{\begin{array}}
\newcommand{\ea}{\end{array}}

\newcommand{\dsR}{{\mathbb R}}

\newcommand{\bm}[1]{\mbox{\boldmath$#1$}}

\def\double #1{#1{\hbox{\kern-2pt $#1$}}}

\newcommand{\bsubeq}{\begin{subequations}}
\newcommand{\esubeq}{\end{subequations}}

\newcommand{\ul}{\underline}
\newcommand{\eps}{{\ve}}

\newcommand{\rd}{\mathrm d}
\newcommand{\veps}{\varepsilon}
\newcommand{\de}{\nabla}
\numberwithin{equation}{section}
\newcommand{\RM}{R(M)}
\newcommand{\RD}{R(\mathbb D)}
\newcommand{\RN}{R(N)}

\newcommand{\RS}{R(S)}
\newcommand{\RK}{R(K)}

\begin{document}
%%%%%%%%%%%%%%%%
%%%%%%%%%%%%%%%%

\begin{titlepage}
\begin{flushright}
Nikhef-2013-018\\
June, 2013\\
\end{flushright}
\vspace{5mm}

\begin{center}
{\Large \bf 
Conformal supergravity in three dimensions: \\
Off-shell actions}
\\ 
\end{center}

\begin{center}

{\bf
Daniel Butter${}^{a}$, Sergei M. Kuzenko${}^{b}$, Joseph Novak${}^{b}$, \\
and
Gabriele Tartaglino-Mazzucchelli${}^{b}$
} \\
\vspace{5mm}

\footnotesize{
${}^{a}${\it Nikhef Theory Group \\
Science Park 105, 1098 XG Amsterdam, The Netherlands}}
~\\
\texttt{dbutter@nikhef.nl}\\
\vspace{2mm}

\footnotesize{
${}^{b}${\it School of Physics M013, The University of Western Australia\\
35 Stirling Highway, Crawley W.A. 6009, Australia}}  
~\\
\texttt{joseph.novak,\,gabriele.tartaglino-mazzucchelli@uwa.edu.au}\\
\vspace{2mm}

\end{center}

\begin{abstract}
\baselineskip=14pt
Using the off-shell formulation for  $\cN$-extended 
conformal supergravity in three dimensions
that has recently been presented in arXiv:1305.3132,  we construct 
superspace actions for conformal supergravity theories with $\cN<6$.
For each of the cases considered, we work out the complete component action 
as well as the gauge transformation laws of the fields belonging to the Weyl supermultiplet. 
The $\cN=1$ and $\cN=2$ component actions derived coincide with those 
proposed by van Nieuwenhuizen and Ro\v{c}ek in the mid-1980s. 
The off-shell $\cN=3$, $\cN=4$ and $\cN = 5$ supergravity actions are new results. 
Upon elimination of the auxiliary fields, these actions reduce
to those constructed by Lindstr\"om and Ro\v{c}ek in 1989 (and also by Gates and Nishino
in 1993).
\end{abstract}

\vfill

\vfill
\end{titlepage}

\newpage
\renewcommand{\thefootnote}{\arabic{footnote}}
\setcounter{footnote}{0}

\tableofcontents

%\newpage

%%%%%%%%%%%%%%%%%%%%%%%%%%%%%%%%%%%%%%%%%%%%%%%%%%%%%%
%%%%%%%%%%%%%%%%%%%%%%%%%%%%%%%%%%%%%%%%%%%%%%%%%%%%%%

\section{Introduction}

In a recent paper \cite{BKNT-M1}, we have developed a new formulation 
for $\cN$-extended conformal supergravity in three dimensions,
which was called conformal superspace.  Inspired by 
the earlier constructed formulations for $\cN=1$ \cite{ButterN=1} and $\cN=2$ \cite{ButterN=2} 
conformal supergravities in four dimensions, it is obtained by gauging 
the $\cN$-extended superconformal algebra in superspace. 
In the framework of \cite{BKNT-M1}, 
the  geometry of curved superspace is subject to covariant constraints  
such that the algebra of covariant derivatives is given in terms of a single 
curvature superfield which turns out to be the super Cotton tensor. 
Upon degauging of the local  $S$-supersymmetry and special conformal transformations, 
the conformal superspace of \cite{BKNT-M1} reduces
to the conventional formulation for $\cN$-extended conformal supergravity \cite{HIPT,KLT-M11}
with the structure group $\rm SL(2 , \dsR) \times SO(\cN)$.
The latter formulation has been used  \cite{KLT-M11} to construct 
 general supergravity-matter couplings in the cases $\cN \leq 4$
(the simplest extended case $\cN=2$ was studied in more detail in \cite{KT-M11}).
However, the approach of  \cite{HIPT,KLT-M11} does not appear to be well suited  
for the explicit construction 
of off-shell conformal supergravity actions in superspace.\footnote{At the component level,
the Chern-Simons type actions for three-dimensional conformal supergravities were  
constructed in the cases  $\cN=1$ \cite{vanN85},  $\cN=2$ \cite{RvanN86}
and finally for arbitrary $\cN$ \cite{LR89,NG}. 
The formulation for extended conformal supergravity given in \cite{LR89,NG} is on-shell for $\cN>2$.}

Within the conventional formulation  \cite{HIPT,KLT-M11},
the superspace action for $\cN=1$ conformal supergravity 
has been worked out in a recent paper \cite{KT-M12}, 
using the ectoplasm formalism \cite{Hasler,Ectoplasm,GGKS},
whose basic principle appeared earlier in the rheonomic
framework \cite{Castellani}.
However, an extension of the general method used in \cite{KT-M12},
which should be applicable in principle for $\cN>1$, 
has faced a formidable amount of calculation already in the case $\cN=2$.  
As discussed  in  \cite{BKNT-M1}, 
the main technical disadvantage of  the conventional formulation  \cite{HIPT,KLT-M11}
is the presence of several dimension-1 curvature tensors  ($S^{IJ} = S^{(IJ)}$,
$C_a{}^{IJ} = C_a{}^{[IJ]}$ and $W^{IJKL} = W^{[IJKL]} $),  which makes the 
algebra of covariant derivatives rather involved and somewhat cumbersome 
from the point of view of practical calculations. 
On the other hand, the conformal superspace of  \cite{BKNT-M1} has no dimension-1 curvature  
for the cases $\cN=1,2,3$, while   for $\cN>3$ the algebra of covariant derivatives
is constructed entirely in terms of the super Cotton tensor $W^{IJKL}$. 

In this paper, we will apply the conformal superspace \cite{BKNT-M1} to construct off-shell 
actions for conformal supergravity theories with $\cN<6$.
In principle, we will make use of the same superform method that was at the heart of the construction 
in \cite{KT-M12} (this method is a generalization of the superform formulation for the linear multiplet in 
four-dimensional $\cN=2$ conformal supergravity given in \cite{BKN12}.) 
However, the technical problems that are intrinsic within the conventional formulation 
\cite{HIPT,KLT-M11} simply do not occur if one works in the conformal superspace of \cite{BKNT-M1}.

This paper is organized as follows. In section \ref{geometry}, we review the recently constructed conformal 
superspace of \cite{BKNT-M1}, emphasizing some key points of the formulation. Section \ref{closedForms} 
presents the necessary framework for the construction of actions from an appropriate closed superform. We 
discuss two three-forms of particular importance, the Chern-Simons and curvature induced three-forms. These 
provide us with an efficient procedure of constructing closed forms, describing the actions in three-dimensions. 
In section \ref{CSCI}, we work out the necessary ingredients for the construction of the supergravity actions -- namely, 
the Chern-Simons and curvature induced three-forms. Section \ref{compAnalysis} is devoted to elaborating the 
component structure of the conformal superspace of \cite{BKNT-M1}. Using the derived component expressions, section 
\ref{actions} constructs the component actions for the $\cN < 6$ cases. Finally, in section \ref{conclusion} 
we discuss our results and conclude the paper.

We also include several technical appendices. Appendix \ref{InvForms} provides some useful notes on 
invariant superforms. In Appendix \ref{matrixReal} we give a quick derivation of the Cartan-Killing metric 
using a matrix realization of the superconformal group.  Appendix \ref{compAnalysis2} includes component analysis of 
certain curvatures, which are not directly used 
for the actions constructed in the paper. We also include supersymmetry transformations of component fields in 
Appendix \ref{SUSY}.

%%%%%%%%%%%%%%%%%%%%%%%%%%%%%%%%%%%%%%%%%%%%%%%%%%%%%
%%%%%%%%%%%%%%%%%%%%%%%%%%%%%%%%%%%%%%%%%%%%%%%%%%%%%

\section{Geometry of $\cN$-extended conformal superspace} \label{geometry}

We begin with a summary of the newly constructed $\cN$-extended conformal
superspace of \cite{BKNT-M1}, which involves gauging the full
superconformal algebra. We then collect the essential details of the
superspace geometry for the distinct cases of $\cN=1$, $\cN=2$, and
$\cN=3$ as well as for $\cN>3$.
These results will be used extensively throughout the rest of the paper.
We refer the reader to appendix A of \cite{BKNT-M1} for our notation and
conventions.

In this section we deal with a curved three-dimensional $\cN$-extended superspace
 $\cM^{3|2 \cN}$ parametrized by
local bosonic $(x^m)$ and fermionic coordinates $(\theta^\m_I)$:
\be z^M = (x^m, \ \q^\mu_I) \ ,
\ee
where $m = 0, 1, 2$, $\mu = 1, 2$ and $I = 1, \cdots , \cN$. 

%%%%%%%%%%%%%%%%%%%%%%%%%%%%%%%%%%%%%%%%%%%%%%%%%%%%%%

\subsection{Gauging the superconformal algebra}

The $\cN$-extended superconformal algebra in three dimensions, 
${\mathfrak{osp}}(\cN|4, {\mathbb R})$,  
can be viewed as an extension of the usual $\cN$-extended super-Poincar\'e algebra,
\begin{subequations} \label{SCA}
\begin{gather}
[M_{ab} , M_{cd}] = 2 \eta_{c[a} M_{b] d} - 2 \eta_{d [a} M_{b] c} \ , \\
[M_{ab} , P_c ] = 2 \eta_{c [a} P_{b]} \ , \\
\{ Q_\a^I, Q_\b^J \} = 2 \ri \,\d^{IJ} (\g^c)_{\a\b} P_c = 2 \ri \,\d^{IJ} P_{\a\b} \ , \\
[M_{\a\b} , Q_\g^I] = \eps_{\g(\a} Q_{\b)}^I \ ,
\end{gather}
(with all other commutators vanishing) by introducing additional bosonic and fermionic 
generators. The extra bosonic generators are the dilatation ($\mathbb D$), special
conformal ($K_a$) and $\rm SO(\cN)$ ($N_{KL}$) generators, while the 
additional fermionic generator is the fermionic special conformal
generator\footnote{The fermionic special conformal operator is also known as the $S$-supersymmetry 
generator.} $(S_\a^I)$.
The purely bosonic part of the extension is given by
\begin{gather}
[M_{ab} , K_c] = 2 \eta_{c[a} K_{b]} \ , \quad  [\mathbb D, K_a] = - K_a \ , \\
[K_a , P_b] = 2 \eta_{ab} \mathbb D + 2 M_{ab} \ , \\
[N_{KL} , N^{IJ}] = 2 \d^I_{[K} N_{L]}{}^J - 2 \d^J_{[K} N_{L]}{}^I \ ,
\end{gather}
while the part involving fermionic generators is
\begin{gather}
[\mathbb D, Q_\a^I] = \hf Q_\a^I \ , \quad [N_{KL} , Q_\a^I] = 2 \d^I_{[K} Q_{\a L]} \ , \\
\{ S_\a^I , S_\b^J \} = 2 \ri \d^{IJ} (\g^c)_{\a\b} K_c \ , \quad [S_\a^I , K_b] = 0 \ , \\
[M_{\a\b} , S_\g^I] = \eps_{\g(\a} S_{\b)}^I \ , \quad [\mathbb D, S_\a^I] = - \hf S_\a^I \ , \quad [N_{KL} , S_\a^I] = 2 \d^I_{[K} S_{\a L]} \ , \\
[K_a , Q_\a^I ] = - \ri (\g_a)_\a{}^\b S_\b^I \ , \quad [S_\a^I , P_a] = \ri (\g_a)_\a{}^\b Q_{\b}^I \ , \\
\{ S_\a^I , Q_\b^J \} = 2 \eps_{\a\b} \d^{IJ} \mathbb D - 2 \d^{IJ} M_{\a \b} - 2 \eps_{\a \b} N^{IJ} \ .
\end{gather}
\end{subequations}
All other (anti-)commutators vanish.

It is convenient to denote the generators of the algebra collectively by $X_{\tilde{a}}$,
with the graded commutators\footnote{The Grassmann parity of $X_{\tilde{a}}$
is denoted by $\eps_{\tilde{a}}$.}
\be [X_{\tilde{a}}, X_{\tilde{b}}\} = - f_{\tilde{a}\tilde{b}}{}^{\tilde{c}} X_{\tilde{c}} \ , \quad f_{\tilde{a}\tilde{b}}{}^{\tilde{c}} = - (-1)^{\eps_{\tilde{a}} \eps_{\tilde{b}}} f_{\tilde{b}\tilde{a}}{}^{\tilde{c}}~.
\ee
The structure constants $f_{\tilde a \tilde b}{}^{\tilde c}$ are graded
anti-symmetric. The algebra satisfies the Jacobi identities
\begin{align} [X_{\tilde{a}}, [X_{\tilde{b}}, X_{\tilde{c}} \} \} + (-1)^{\eps_{\tilde{a}} (\eps_{\tilde{b}} + \eps_{\tilde{c}})} [X_{\tilde{b}}, [X_{\tilde{c}}, X_{\tilde{a}} \} \}
+ (-1)^{\eps_{\tilde{c}} (\eps_{\tilde{a}} + \eps_{\tilde{b}})} [X_{\tilde{c}} , [ X_{\tilde{a}} , X_{\tilde{b}} \} \} = 0 \ ,
\end{align}
which can be compactly written as
\be f_{[\tilde{a}\tilde{b}}{}^{\tilde{d}} f_{|\tilde{d}| \tilde{c} \} }{}^{\tilde{e}} = 0 \ . \label{SCJI}
\ee

In order to gauge this algebra, we must draw a distinction between the
generators $P_A = (P_a, Q_\alpha^I)$ and the remaining
generators $X_{\ul a}$, which generate a subgroup $\cH$ of the superconformal group.
We introduce the vielbein one-form $E^A = \rd z^M E_M{}^A$ and the additional
one-forms $\omega^{\underline{a}} = \rd z^M \omega_M{}^{\underline{a}}$,
and we require them to transform under $\cH$ as
\begin{align}\label{eq:deltaHConn}
\delta_\cH E^A = E^{C} \L^{\ul b} f_{\ul b C}{}^A~, \qquad
\delta_\cH \omega^{\ul a} = \rd \L^{\ul a} + E^{C} \L^{\ul b} f_{\ul b C}{}^{\ul a}
	+ \omega^{\ul c} \L^{\ul b} f_{\ul b \ul c}{}^{\ul a}~.
\end{align}
From these connections, we construct the covariant derivative
\be
\nabla_A = E_A{}^M \nabla_M = E_A{}^M \partial_M - \omega_A{}^{\underline{a}} X_{\underline{a}} \ , \qquad \omega_A{}^{\underline{a}} = E_A{}^M \omega_M{}^{\underline{a}} \ ,
\ee
where $E_A{}^M$ is the inverse vielbein.
As a consequence of eq. \eqref{eq:deltaHConn}, the action of $X_{\underline{a}}$ on $\nabla_A$ mimics that of $X_{\ul a}$ on $P_A$,
\be [X_{\underline{a}} , \nabla_A \} = - f_{\underline{a} P_A}{}^{P_B} \nabla_B - f_{\underline{a} P_A}{}^{\underline{b}} X_{\underline{b}} \ .
\ee
Therefore, the covariant derivatives $\nabla_A$ together with the generators $X_{\ul a}$
satisfy nearly the same algebra \eqref{SCA} we started with (replacing $P_A \rightarrow \nabla_A$),
except for the introduction of torsion and curvatures,
\begin{align}
[\nabla_A , \nabla_B \} = - T_{AB}{}^C \nabla_C - R_{AB}{}^{\underline{c}} X_{\underline{c}} ~,
\end{align}
which are given respectively by
\begin{subequations}\label{eq:defTR}
\begin{align}
T^A &:= \hf E^C \wedge E^B T_{BC}{}^A = \rd E^A - E^C \wedge \omega^{\underline{b}} \,f_{\underline{b} C}{}^A \ , \\
R^{\underline{a}} &:= \hf E^C \wedge E^B R_{BC}{}^{\underline{a}} = \rd \omega^{\underline{a}}
	- E^C \wedge \omega^{\underline{b}} \, f_{\underline{b} C}{}^{\underline{a}}
	- \hf \omega^{\underline{c}} \wedge \omega^{\underline{b}} \,
		f_{\underline{b} \underline{c}}{}^{\underline{a}} \ .
\end{align}
\end{subequations}
In particular, the new algebra obeys the Jacobi identities
\be [X_{\underline{a}} , [ \nabla_B , \nabla_C \}\} + {\rm permutations} = 0 \ ,\label{eq:Jacobi}
\ee
which determine the gauge transformations of the torsion and curvatures, and
\be [\nabla_A , [\nabla_B , \nabla_C \} \} + {\rm permutations} = 0 \ , \label{CDBI}
\ee
which are equivalent to the usual Bianchi identities.

The full gauge group  of conformal supergravity, $\cG$, consists of 
{\it covariant general coordinate transformations}, 
$\delta_{\rm cgct}$, associated with a parameter $\xi^A$ and 
{\it standard superconformal 
transformations},\footnote{This terminology follows \cite{FVP}.} 
$\delta_{\cH}$, denoted by a parameter $\L^{\ul a}$. 
The latter include
the dilatations,
Lorentz transformations, $\rm SO(\cN)$ transformations, and special conformal
(bosonic and fermionic) transformations.
The covariant derivatives transform as
\bea 
\d_\cG \nabla_A &=& [\cK , \nabla_A] \ ,
\label{TransCD}
\eea
where $\cK$ denotes the first-order differential operator
\bea
\cK = \xi^C \nabla_C + \L^{\underline{a}} X_{\underline{a}}~.
\eea
Covariant (or tensor) superfields transform as
\bea 
\d_{\cG} T &=& \cK T~.
\eea
The transformations of the one-forms $E^A$ and $\omega^{\ul a}$ are
\begin{subequations}\label{eq:deltaGConn}
\begin{align}
\delta_\cG E^A &= \rd \xi^A
	+ \omega^{\ul c} \xi^{B} f_{B \ul c}{}^A
	+ E^C \xi^B T_{BC}{}^A
	+ E^{C} \L^{\ul b} f_{\ul b C}{}^A~, \\
\delta_\cG \omega^{\ul a} &= \rd \L^{\ul a} + E^{C} \L^{\ul b} f_{\ul b C}{}^{\ul a}
	+ \omega^{\ul c} \L^{\ul b} f_{\ul b \ul c}{}^{\ul a}
	+ \omega^{\ul c} \xi^{B} f_{B \ul c}{}^{\ul a}
	+ E^C \xi^B R_{BC}{}^{\ul a} ~,
\end{align}
\end{subequations}
which are equivalent to eq. \eqref{TransCD}.

It is important to note that we have not gauged the algebra in the conventional
sense.\footnote{For pedagogical reviews of superconformal supergravity in four dimensions, 
see e.g. \cite{VanN81,FradTsey,FVP}.}  
If we had treated the vielbein $E^A$ and the connections $\omega^{\ul a}$
on a completely equivalent footing, we could have introduced the notation
\be \omega^{\tilde{a}} = (E^A , \omega^{\underline{a}}) ~,
\ee
and postulated the usual gauge transformations 
\be 
\d_\L \omega^{\tilde{a}} = \rd \L^{\tilde{a}} 
+ \omega^{\tilde{c}} \L^{\tilde{b}} f_{\tilde{b} \tilde{c}}{}^{\tilde{a}} \ , \quad \L^{\tilde{a}} = (\zeta^A , \L^{\underline{a}}) \ .
\label{TConnec}
\ee
These coincide with eq. \eqref{eq:deltaGConn} for pure $\cH$-transformations
$\L^{\tilde a} = (0,\L^{\ul a})$, but the $P$-gauge transformations
(with parameter $\zeta^A$) do \emph{not} coincide with covariant
diffeomorphisms (with parameter $\xi^A$) except when the curvature
$R^{\ul a}$ vanishes and the torsion $T_{BC}{}^A$ takes the constant
value $f_{BC}{}^A$.

Nevertheless, it can be extremely advantageous to group our connections
together in this way. The main reason will be that the conventionally-defined
curvatures of a gauge theory
\begin{align}
R^{\tilde{a}} := \rd \omega^{\tilde{a}} 
- \hf \omega^{\tilde{c}} \wedge \omega^{\tilde{b}} f_{\tilde{b} \tilde{c}}{}^{\tilde{a}}
\end{align}
are nearly identical to the \emph{actual} curvatures defined in eq. \eqref{eq:defTR}.
The $\cH$-curvatures $R^{\ul a}$ are precisely the same, while the
curvature $R(P)^A$ associated with $P_A$ is simply the difference between
the curved and flat torsions,
\be
R(P)^{A} = T^A - \hf E^C \wedge E^B f_{B C}{}^{A} ~.
\ee
Moreover, the curvatures $R^{\tilde a} = (R(P)^A, R^{\ul a})$ vanish precisely when the
superspace is flat.
One can further show that the curvatures $R^{\tilde a}$ transform under the $\cH$-gauge
transformations as
\be 
\d_\cH R^{\tilde{a}} = R^{\tilde{c}} \L^{\tilde{b}} f_{\tilde{b} \tilde{c}}{}^{\tilde{a}} ~,
\qquad \L^{\tilde b} = (0, \L^{\ul b}) \label{TCurv}
\ee
and obey the Bianchi identities
\be \rd R^{\tilde{a}} + R^{\tilde{c}} \wedge \omega^{\tilde{b}} f_{\tilde{b} \tilde{c}}{}^{\tilde{a}} = 0~. \label{BIcurv}
\ee
These conditions are equivalent, respectively, to eqs. \eqref{eq:Jacobi} and \eqref{CDBI},
and will be extremely useful for our later construction of the Chern-Simons action.

%%%%%%%%%%%%%%%%%%%%%%%%%%%%%%%%%%%%%%%%%%%%%%%%%%%%%%

\subsection{$\cN$-extended conformal supergravity}

Specializing now to $\cN$-extended conformal supergravity, the covariant derivatives
have the form
\be
\nabla_A = E_A{}^M \pa_M - \o_A{}^{\underline b} X_{\underline b} 
= E_A{}^M \pa_M - \hf \Omega_A{}^{ab} M_{ab} - \hf \Phi_A{}^{PQ} N_{PQ} - B_A \mathbb D - \mathfrak{F}_A{}^B K_B \ ,
\ee
and satisfy the (anti-)commutation relations
\begin{align}
[ \nabla_A , \nabla_B \}
	&= -T_{AB}{}^C \nabla_C
	- \frac{1}{2} \RM_{AB}{}^{cd} M_{cd}
	- \frac{1}{2} \RN_{AB}{}^{PQ} N_{PQ}
	\non \\ & \quad
	- \RD_{AB} \mathbb D
	- \RS_{AB}{}^\g_K S_\g^K
	- \RK_{AB}{}^c K_c~.
\end{align}
The explicit expressions for the torsion and various curvatures follow from eq. \eqref{eq:defTR}.
We give them in their entirety here for later reference:
\begin{subequations}\label{eq:TRcsg}
\bea
T^a &=& \rd E^a + E^a \wedge B + E^b \wedge \Omega_b{}^a \ ,
	\label{Tor1} \\
T^\a_I &=& \rd E^\a_I + \hf E^\b_I \wedge \Omega^{c} (\g_c)_\b{}^\a + \hf E^\a_I \wedge B + E^{\a J} \wedge \Phi_{JI} + \ri \, E^c \wedge \mathfrak{F}^\beta_I (\gamma_c)_\beta{}^\alpha \ ,~~~~~~~~~~~ 
	\label{Tor2} \\
\RD &=& \rd B + 2 E^a \wedge \mathfrak{F}_a - 2 E^\a_I \wedge \mathfrak{F}_\a^I \ ,
	\label{Dcurv} \\
\RM^{ab} &=& \rd \Omega^{ab} + \Omega^{ac} \wedge \Omega_c{}^b - 4 E^{[a} \wedge \mathfrak{F}^{b]} - 2 E^\a_I \wedge \mathfrak{F}^{\b I} (\g_c)_{\a\b} \eps^{cab} \ ,
	\label{MCurv} \\
\RN^{IJ} &=& \rd \Phi^{IJ} + \Phi^{I K} \wedge \Phi_{K}{}^J - 4 E^{\a [I} \wedge \mathfrak{F}_{\a}{}^{J]} \ ,
	\label{NCurv} \\
\RK^a &=& \rd \mathfrak{F}^a - \mathfrak{F}^a \wedge B + \mathfrak{F}^b \wedge \Omega_b{}^a + \ri \mathfrak{F}^\a_I \wedge \mathfrak{F}^{\b I} (\g^a)_{\a\b} \ ,
	\label{KCurv} \\
\RS^\a_I &=& \rd \mathfrak{F}^\a_I - \ri E^\b_I \wedge \mathfrak{F}^a (\g_a)_\b{}^\a - \hf \mathfrak{F}^{\a}_I \wedge B + \hf \mathfrak{F}^\b_I \wedge \Omega^{c} (\g_c)_\b{}^\a + \mathfrak{F}^{\a J} \wedge \Phi_{JI}
	\label{SCurv} \ .
\eea
\end{subequations}

In order to describe conformal supergravity irreducibly in superspace,
it is necessary to constrain the above torsion and curvature tensors.
The appropriate constraints  were given in \cite{BKNT-M1} and were based on two principles.
First, the curvatures should be expressed in terms of a single conformal primary
superfield, the $\cN$-extended super Cotton tensor.\footnote{The super Cotton tensors for $\cN = 1$,
$\cN = 2$ and $\cN=3$ are described by superfields $W_{\a\b\g} = W_{(\a\b\g)}$,
 $W_{\a\b} = W_{(\a\b)}$ and  $W_\a$, which were given in  
\cite{KT-M12},  \cite{ZP,Kuzenko12} and  \cite{BKNT-M1} respectively. 
For $\cN > 3$  the super Cotton tensor is  
a totally antisymmetric $\rm SO(\cN)$ superfield $W^{IJKL} = W^{[IJKL]}$ \cite{HIPT}.} 
Second, the constraints imposed on the geometry should resemble super Yang-Mills.
These two basic principles turn out to uniquely determine the torsion and
curvatures for all values of $\cN$.

Before giving the algebra of covariant derivatives, we should point out
one important feature which is independent of the choice of $\cN$. The torsion
tensor always takes its constant flat space value, while the Lorentz and dilatation
curvatures always vanish:\footnote{These constraints appear to be the superspace analogue of those in \cite{vanN85, RvanN86}.}
\begin{subequations}
\begin{align}
T^a &= -\ri \, E^\beta \wedge E^\gamma (\gamma^a)_{\gamma \beta}~, \quad T^\alpha_I = 0 \qquad
\Longleftrightarrow  \qquad R(P)^A = 0   \label{RPConstraint}~;
\\
\RM^{ab} &= 0~, \qquad \RD = 0  ~. \label{RMDConstraint}
\end{align}
\end{subequations}
We now give the covariant derivative algebra, including the explicit form
for the remaining curvatures, for all values of $\cN$.

%%%%%%%%%%%%%%%%%%%%%%%%%%%%%%%%%%%%%%%%%%%%%%%%%%%%%%

\subsubsection{The $\cN = 1$ case}

The $\cN=1$ super Cotton tensor $W_{\a\b\g}$ is a symmetric primary superfield of dimension-$5/2$
\be S_\d W_{\a\b\g} = 0 \ , \quad \mathbb D W_{\a\b\g} = \frac{5}{2} W_{\a\b\g} ~.
\ee
The algebra of covariant derivatives is given by
\begin{subequations}
\begin{align} \{ \nabla_\a , \nabla_\b \} &= 2 \ri \nabla_{\a\b} \ , \\
[ \nabla_a , \nabla_\a ] &= \frac{1}{4} (\g_a)_\a{}^\b W_{\b \g\d} K^{\g\d} \ , \\
[\nabla_a , \nabla_b] &= - \frac{\ri}{8} \eps_{abc} (\g^c)^{\a\b} \nabla_\a W_{\b\g\d} K^{\g\d} - \frac{1}{4} \eps_{abc} (\g^c)^{\a\b} W_{\a\b\g} S^\g \ .
\end{align}
\end{subequations}
The Bianchi identities \eqref{CDBI} imply an additional constraint on $W_{\a\b\g}$:
its spinor divergence must vanish,
\be \nabla^\a W_{\a \b\g} = 0 \ .
\ee

%%%%%%%%%%%%%%%%%%%%%%%%%%%%%%%%%%%%%%%%%%%%%%%%%%%%%%

\subsubsection{The $\cN = 2 $ case}

The $\cN = 2$ super Cotton tensor $W_{\a\b}$ is a symmetric primary superfield of dimension-$2$
\be S_\g^I W_{\a\b} = 0 \ , \quad \mathbb D W_{\a\b} = 2 W_{\a\b} ~.
\ee
As in the $\cN=1$ case, its spinor divergence vanishes,
\be \nabla^{\a I} W_{\a\b} = 0 \ .
\ee
The algebra of covariant derivatives is 
\begin{subequations} \label{N=2Algebra}
\begin{align} \{ \nabla_\a^I , \nabla_\b^J \} &= 2 \ri \d^{IJ} \nabla_{\a\b} - \ri \eps^{IJ} \eps_{\a\b} W_{\g\d} K^{\g\d} \ , \\
[ \nabla_a , \nabla_\b^J ] &= \hf (\g_a)_\b{}^\g \eps^{JK} \nabla_{\g K} W^{\a\d} K_{\a\d} + \ri (\g_a)_{\b\g} \eps^{JK} W^{\g\d} S_{\d K} \ , \\
[\nabla_a , \nabla_b] &= - \frac{\ri}{8} \eps_{abc} (\g^c)^{\g\d} \Big( \eps^{KL} (\nabla_{\g K} \nabla_{\d L} W_{\a\b} K^{\a\b} + 4 \ri \nabla_{\g K} W_{\d\b} S^\b_L) - 8 W_{\g\d} \cJ \Big) \ , 
\end{align}
\end{subequations}
where the $\rm U(1)$ generator $\cJ$ obeys
\be N_{KL} = \ri \eps_{KL} \cJ \ , \quad \cJ = - \frac{\ri}{2} \eps^{KL} N_{KL} ~, \qquad
[\cJ , \nabla_\a^I] = - \ri \eps^{IJ} \nabla_{\a J} \ .
\ee

%%%%%%%%%%%%%%%%%%%%%%%%%%%%%%%%%%%%%%%%%%%%%%%%%%%%%%

\subsubsection{The $\cN = 3$ case}

The $\cN = 3$ super Cotton tensor $W_{\a}$ is a primary superfield of dimension-$3/2$
with vanishing spinor divergence,
\be S_\b^I W_{\a} = 0 \ , \qquad \mathbb D W_{\a} = \frac{3}{2} W_{\a}~, \qquad
\nabla^{\alpha I} W_\alpha = 0~.
\ee
The algebra of covariant derivatives is
\begin{subequations} \label{N=3Algebra}
\begin{align} 
\{ \nabla_\a^I , \nabla_\b^J \} &= 2 \ri \d^{IJ} \nabla_{\a\b} - 2 \eps_{\a\b} \eps^{IJL} W^\g S_{\g L} + \ri \eps_{\a\b} (\g^c)^{\g\d} \eps^{IJK} (\nabla_{\g K} W_\d) K_c \ , \\
 [\nabla_a , \nabla_\b^J ] &= \ri \eps^{JKL} (\g_a)_{\b\g} W^\g N_{KL} + \ri \eps^{JKL} (\g_a)_{\b\g} (\nabla^\g_K W^\d) S_{\d L} \non\\
 &\qquad + \frac{1}{4} \eps^{JKL} (\g_a)_{\b\g} (\g^c)_{\d\rho} (\nabla^\g_K \nabla^\d_L W^\rho) K_c \ , \\
[\nabla_a , \nabla_b] &= \eps_{abc} (\g^c)_{\a\b} \Big[ - \hf \eps^{IJK} (\nabla^\a_I W^\b) N_{JK} - \frac{1}{4} \eps^{IJK} (\nabla^\a_I \nabla^\b_J W^\g ) S_{\g K} \non\\
&\qquad +\frac{\ri}{24} \eps^{IJK} (\g^d)_{\g\d} (\nabla^\a_I \nabla^\b_J \nabla^\g_K W^\d) K_d \Big] \ .
\end{align}
\end{subequations}

%%%%%%%%%%%%%%%%%%%%%%%%%%%%%%%%%%%%%%%%%%%%%%%%%%%%%%

\subsubsection{The $\cN > 3$ case}

For all values of $\cN>3$, we introduce the super Cotton tensor $W^{IJKL}$,
which is a totally antisymmetric primary superfield of dimension-1
\be
S_\alpha^P W^{IJKL} = 0~, \quad \mathbb D W^{IJKL} = W^{IJKL} \ .
\ee
The algebra of covariant derivatives is\footnote{The algebra for $\cN \leq 3$ can be deduced 
from that for $\cN > 3$ \cite{BKNT-M1}.}
\begin{subequations} \label{covDN>3}
\begin{align} 
\{ \nabla_\a^I , \nabla_\b^J \} &= 2 \ri \d^{IJ} \nabla_{\a\b} + \ri \eps_{\a\b} W^{IJKL} N_{KL} - \frac{\ri}{\cN - 3} \eps_{\a\b} (\nabla^\g_K W^{IJKL}) S_{\g L} \non\\
&\qquad + \frac{1}{2 (\cN - 2)(\cN - 3)} \eps_{\a\b} (\g^c)^{\g\d} (\nabla_{\g K} \nabla_{\d L} W^{IJKL}) K_c \ , \\
[\nabla_a , \nabla_\b^J ] &= \frac{1}{2 (\cN - 3)} (\g_a)_{\b\g} (\nabla^\g_K W^{JPQK}) N_{PQ} \non\\
&\qquad - \frac{1}{2 (\cN - 2) (\cN - 3)} (\g_a)_{\b\g} (\nabla^\g_L \nabla^\d_P W^{JKLP}) S_{\d K} \non\\
&\qquad - \frac{\ri}{4 (\cN - 1) (\cN - 2)(\cN - 3)} (\g_a)_{\b\g} (\g^c)_{\d\rho} (\nabla^\g_K \nabla^\d_L \nabla^\rho_P W^{JKLP}) K_c \ , \\
[\nabla_a , \nabla_b] &
=    \frac{1}{4 (\cN - 2) (\cN - 3)}   \eps_{abc} (\g^c)_{\a\b} 
\Big( \ri (\nabla^\a_I \nabla^\b_J W^{PQIJ}) N_{PQ} \non\\
&\qquad + \frac{\ri}{ \cN - 1} (\nabla^\a_I \nabla^\b_J \nabla^\g_K W^{LIJK}) S_{\g L} \non\\
&\qquad + \frac{1}{2 \cN (\cN - 1) } (\g^d)_{\g\d} (\nabla^\a_I \nabla^\b_J \nabla^\g_K \nabla^\d_L W^{IJKL}) K_d \Big) \ ,
\end{align}
\end{subequations}
where $W^{IJKL}$ satisfies the Bianchi identity
\be \nabla_{\a}^I W^{JKLP} = \nabla_\a^{[I} W^{JKLP]} - \frac{4}{\cN - 3} \nabla_{\a Q} W^{Q [JKL} \d^{P] I} \ . \label{CSBIN>3}
\ee

For $\cN=4$, the equation eq. \eqref{CSBIN>3} is trivially satisfied,
and instead  a fundamental Bianchi identity occurs at  dimension-2. 
Rewriting the super Cotton tensor as a scalar superfield,
$W^{IJKL}:=\ve^{IJKL}W$, the Bianchi identity reads
\bea
\nabla^{\a I}\nabla_{\a}^JW=\frac{1}{4}\d^{IJ}\nabla^{\a}_P\nabla_{\a}^PW~. \label{2.39}
\eea

For further details about this superspace formulation, we refer
the reader to \cite{BKNT-M1}.

%%%%%%%%%%%%%%%%%%%%%%%%%%%%%%%%%%%%%%%%%%%%%%%%%%%%%
%%%%%%%%%%%%%%%%%%%%%%%%%%%%%%%%%%%%%%%%%%%%%%%%%%%%%

\section{Closed three-forms and locally superconformal actions} \label{closedForms}

Traditionally, the supersymmetric actions were (and in many cases still are) 
realized as integrals over the full superspace
or its invariant subspaces. A paradigm shift took place in the late 1990s 
when the superform (or ectoplasm) approach 
for the construction  of supersymmetric invariants 
\cite{Hasler,Ectoplasm,GGKS} was introduced.\footnote{The ectoplasm approach has become
a powerful tool for the construction and analysis of counterterms in extended supergravity theories, 
see \cite{BHLSW,BHS13} and references therein.
%An identical method for the construction of supersymmetric
%Lagrangians was used earlier in 
The ectoplasm approach proves to be equivalent 
to the rheonomic formalism
\cite{Castellani} which was developed several years earlier. 
Unfortunately,  the latter approach remained unknown to many 
superspace practitioners. }
In the case of three-dimensional spacetime $\cM^3$, which is the body 
of the $\cN$-extended curved superspace $\cM^{3|2\cN}$,
the formalism requires the use of 
a closed three-form
\bea
\frak{J} = \frac{1}{3 !} E^C \wedge E^B \wedge E^A \frak{J}_{ABC} 
=  \frac{1}{3 !} \rd z^R \wedge \rd z^N \wedge \rd z^M \frak{J}_{MNR} 
\ ,\qquad
 \rd \frak{J} = 0 ~. 
\eea
Given such a superform, 
it is a short calculation to show 
that the action\footnote{The Levi-Civita tensor with world indices 
is defined as $\eps^{mnp} := \eps^{abc} e_a{}^m e_b{}^n e_c{}^p$.}
\bea 
S = \int_{\cM^3} \frak{J} = \int \rd^3 x \,e \,{}^* \frak{J} |_{\q=0}\ , \qquad 
{}^*\frak{J} = \frac{1}{3!} \eps^{mnp} \frak{J}_{mnp} 
\label{ectoS}
\eea
is invariant under arbitrary general coordinate transformations of the superspace. 
To see this we only need to re-iterate the proof given by Hasler 
in four dimensions \cite{Hasler} (see also \cite{GGKS}). 
Under an infinitesimal coordinate transformation (or diffeomorphism)
generated by a vector field $\x = \x^A E_A = \x^M \pa_M$, 
the three-form varies as  
\be 
\d_{\xi} \frak{J} = \cL_\xi \frak{J} \equiv i_{\xi} \rd \frak{J} + \rd i_{\xi} \frak{J} 
= \rd i_{\xi} \frak{J} \ . 
\ee
Since the variation $\d_{\xi} \frak{J}$ is an exact form, the action $S$ is indeed invariant under general coordinate transformations
provided  the components $\x^M$ vanish at the boundary of  
the spacetime $\cM^3$.

In $\cN$-extended conformal supergravity,  suitable  actions  
must also be  invariant under the {\it standard superconformal 
transformations}.
 If the closed three-form $\frak{J}$ also transforms by an exact form 
 under the standard superconformal  transformations, 
\bea
 \d_{\cH} \frak{J} =  \rd \Theta (\L^{\underline{a}} ) \ , \quad 
\L = \L^{\underline{a}} X_{\underline{a}}~,
\eea
then the functional \eqref{ectoS} is a suitable candidate for an action.
The explicit structure of the two-form $\Theta (\L^{\underline{a}} ) $ 
is constrained due to
the fact that
the standard superconformal  transformations form a closed algebra. 
As will be shown below, the conformal supergravity actions  
with $\cN<6$ 
provide examples of closed three-forms with a non-zero $\Q$.
As concerns locally superconformal matter actions, 
in most cases they are associated 
with closed invariant three-forms such that 
\bea
\d_{\cH} \frak{J} =  0~.
\label{3.5}
\eea
This is expected to be the case for the super-Weyl invariant action functionals 
constructed in \cite{KLT-M11} for the cases  $\cN<5$.
Implications of the invariance condition \eqref{3.5} are spelled out in Appendix 
 \ref{InvForms}.

As an example of a closed invariant three-form, we choose $\cN=1$ 
and lift  the super-Weyl invariant three-form constructed in \cite{KT-M12} 
to the conformal superspace.\footnote{The three-form \eqref{3.5} was originally introduced in \cite{BCGLMP}. However, the 
authors of \cite{BCGLMP} did not notice super Weyl invariance of the form.} The result is 
\bea
\X =
\frac{\ri}{2}E^\g\wedge E^\b\wedge E^a
(\g_a)_{\b\g}\cL
&+&\frac{1}{4}E^\g\wedge E^b\wedge E^a\ve_{abc}
(\g^c)_{\g}{}^\d\nabla_{\d}\cL
\non\\
&-&\frac{\ri}{24}E^c\wedge E^b\wedge E^a\ve_{abc}
\nabla^\d \nabla_\d  \cL~,
\label{2.21}
\eea
where the real scalar $\cL$ is a primary superfield of dimension-$2$. 
It is an instructive exercise to check explicitly that this form is closed, $\rd \,\X =0$, 
and obeys the invariance condition \eqref{3.5}. 

The appropriate closed three-form $\frak{J}$ for the action of conformal supergravity 
may be found with the use of two specific three-forms. 
These are the Chern-Simons and curvature induced three-forms.\footnote{The 
approach here is a generalization of the method proposed in  \cite{KT-M12}, 
which in turn is a generalization of the ectoplasm formulation for the linear multiplet in four-dimensional $\cN = 2$ conformal supergravity \cite{BKN12}.}

To construct the Chern-Simons three-form, we have to make use of 
a non-degenerate Cartan-Killing metric of the $\cN$-extended 
superconformal algebra \eqref{SCA}. The  Cartan-Killing metric\footnote{For certain 
special cases this expression may vanish, in which case the fundamental representation must be used to define the metric, see Appendix \ref{matrixReal}.}
can be defined in terms of the structure constants as 
\be 
\G_{\tilde{a} \tilde{b}} = f_{\tilde{a} \tilde{d}}{}^{\tilde{c}} f_{\tilde{b} \tilde{c}}{}^{\tilde{d}} (-1)^{\eps_{\tilde{c}}} \ . 
\label{KillingM} 
\ee
It possesses the following algebraic properties
\begin{subequations}
\begin{align} \G_{\tilde{a} \tilde{b}} &= (-1)^{\eps_{\tilde{a}} \eps_{\tilde{b}}} \G_{\tilde{b} \tilde{a}} \ , \label{KM-sym} \\
f_{\tilde{a} \tilde{b}}{}^{\tilde{d}} \G_{\tilde{d} \tilde{c}} &= - (-1)^{\eps_{\tilde{b}} \eps_{\tilde{c}}} f_{\tilde{a} \tilde{c}}{}^{\tilde{d}} \G_{\tilde{d} \tilde{b}} \ . \label{KM-invariant}
\end{align}
\end{subequations}
${}$From the above relations we see that the structure constants with all indices lowered
\be f_{\tilde{a} \tilde{b} \tilde{c}}  := f_{\tilde{a} \tilde{b}}{}^{\tilde{d}} \G_{\tilde{d} \tilde{c}} 
\ee
are graded antisymmetric.

Using the Cartan-Killing metric we can construct a gauge invariant closed four-form
\be \langle R^2 \rangle := R^{\tilde{b}} \wedge R^{\tilde{a}} \G_{\tilde{a} \tilde{b}} \ , \quad \rd \langle R^2 \rangle = 0 \ .
\ee
The superform $\langle R^2 \rangle$ is $\cH$-gauge invariant\footnote{Keep in mind that a covariant general coordinate transformation $\d_{\rm cgct}$ 
is a combination of a coordinate transformation and a special choice of $\cH$-gauge transformation.}
\be \d_\cH \langle R^2 \rangle = 0 \ ,
\ee
by virtue of the eqs. \eqref{TCurv} and \eqref{KM-invariant}, while its closure
\be \rd \langle R^2 \rangle = 0
\ee
is the result of the eqs. \eqref{BIcurv} and \eqref{KM-invariant}. 
Extracting a total exterior 
derivative from 
$\langle R^2 \rangle$ gives us the Chern-Simons three-form
\bea 
\S_{\rm CS} = R^{\tilde{b}} \wedge \omega^{\tilde{a}} \G_{\tilde{a} \tilde{b}} 
+ \frac{1}{6} \omega^{\tilde{c}} \wedge \omega^{\tilde{b}} \wedge \omega^{\tilde{a}} f_{\tilde{a} \tilde{b} \tilde{c}} \ , \quad \rd \S_{\rm CS} = \langle R^2 \rangle \ . 
\label{CSform}
\eea
Since $\S_{\rm CS}$
has been constructed by extracting
a total exterior derivative
from 
$\langle R^2 \rangle$ it can only transform by a closed form 
under the standard superconformal transformations. 
In fact, it transforms by an exact form under the $\cH$-gauge group
\be \d_{\cH} \S_{\rm CS} = \rd (\rd \omega^{\tilde{b}} \L^{\tilde{a}} \G_{\tilde{a}\tilde{b}}) \ , \qquad
\L^{\tilde a} = (0, \L^{\ul a})~.
\ee

If the off-shell action for conformal supergravity comes from a closed three-form 
$ \frak{J}$, the old component results \cite{vanN85,RvanN86,LR89} 
tell us  that a part of $\frak J$ should be the Chern-Simons three-form.
Then we must have
\be \frak{J} = \S_{\rm CS} - \S_R \ ,
\ee
where $\S_R = \frac{1}{3 !} E^{C} \wedge E^{B} \wedge E^{A} 
\S_{A  B C}  $ is another solution to the superform equation
\be \rd \S_R = \langle R^2 \rangle~.
\ee
In some cases $\langle R^2 \rangle$ proves to vanish 
(as will be discussed below, this is actually true for $\cN=1$ and $\cN=2$)
and then the Chern-Simons three-form is closed automatically, 
$ \rd \S_{\rm CS} =0$, and thus $\S_R = 0$.
If however $\langle R^2 \rangle \neq 0$, then 
$\S_R$ 
is expected to be an invariant three-form, 
\be 
\d_\cH \S_R = 0 \ .
\ee
In accordance with the analysis in Appendix \ref{InvForms}, this implies that 
(i) $\S_{A B C}$ is a tensor under the local Lorentz and SO$(\cN)$ groups; 
and (ii)  the lowest (by dimension) non-zero component of $\S_{A B C}$ is a primary 
superfield. These results mean that $\S_R$ is constructed in terms of the super Cotton tensor
and its covariant derivatives. 
We will refer to $\S_R$ as the curvature induced three-form.\footnote{This was
called the torsion induced three-form and denoted by $\S_T$ in \cite{KT-M12}.} 
The curvature induced form 
appears to exist only for special values of $\cN$.

%%%%%%%%%%%%%%%%%%%%%%%%%%%%%%%%%%%%%%%%%%%%%%%%%%%%%%

\section{Conformal supergravity actions}  \label{CSCI}

In this section, we first elaborate on the Chern-Simons form for general $\cN$ and then construct the 
curvature induced forms for $\cN < 6$. The resulting three-forms will live on the
full superspace $\cM^{3|2 \cN}$. We address their restriction to the bosonic
manifold $\cM^3$ and the explicit construction of the corresponding component
actions in section \ref{actions}.

\subsection{The Chern-Simons three-form}\label{subsec4.1}

In order to compute $\langle R^2 \rangle$ and the Chern-Simons form, it
is necessary to evaluate the components of the Cartan-Killing metric
$\G_{\tilde{a} \tilde{b}}$. One finds that the only non-zero 
components of the Cartan-Killing metric are
\begin{gather}
\G_{M_{a b} M_{c d}} \ , \quad \G_{P_a K_b} = \G_{K_b P_a} \ , \quad
	\G_{\mathbb D \mathbb D} \ , \quad \G_{Q_\a^I S_\b^J} = - \G_{S_\b^J Q_\a^I} \ , \quad
	\G_{N_{IJ} N_{KL}} \ .
\end{gather}
They may be computed directly from the definition \eqref{KillingM}. One finds\footnote{The
 the Cartan-Killing metric vanishes for $\cN=6$ in the adjoint representation
 but it remains non-degenerate for any $\cN$  
 in the fundamental representation, see Appendix \ref{matrixReal}.
 More generally, the Cartan-Killing metric of the superalgebra $\frak{osp}(n|2m,{\mathbb R})$, 
 with $n$ and $m$ positive integers,
 vanishes for $n-2m =2$ in the adjoint representation \cite{Kac}, 
 see \cite{DeWitt} for a review.}
 \bsubeq \label{KillingComps}
\be \G_{\tilde{a} \tilde{b}} = (\cN - 6) {\bm \G}_{\tilde{a} \tilde{b}} \ ,
\ee
where
\begin{align}
{\bm \G}_{M_{a b} M_{c d}} &= 2 \eta_{a [c} \eta_{d] b} \ , \quad
	{\bm \G}_{K_b P_a} = 2 \eta_{ab} \ , \quad {\bm \G}_{\mathbb D \mathbb D} = -1 \ , \\
{\bm \G}_{Q_\a^I S_\b^J} &= 4 \d^{IJ} \eps_{\a\b}\ , \quad
	{\bm \G}_{N_{IJ} N_{KL}} = - 4 \d_{K [I} \d_{J]L } \ .
\end{align}
\esubeq
These components may be compared with those derived in \cite{vanN85, RvanN86, NG}
(see also Appendix \ref{matrixReal}).

To avoid awkward factors of $\cN - 6$, it is convenient to introduce the renormalized
Chern-Simons form
\be {\bm \S}_{\rm CS} = \frac{1}{\cN - 6} \S_{\rm CS} = R^{\tilde{b}} \wedge \omega^{\tilde{a}} {\bm \G}_{\tilde{a} \tilde{b}} 
+ \frac{1}{6} \omega^{\tilde{c}} \wedge \omega^{\tilde{b}} \wedge \omega^{\tilde{a}} {\bm f}_{\tilde{a}\tilde{b}\tilde{c}} \ , \label{eq:SigCS}
\ee
where ${\bm f}_{\tilde{a}\tilde{b}\tilde{c}} =  f_{\tilde{a}\tilde{b}}{}^{\tilde{d}} {\bm \G}_{\tilde{d} \tilde{c}}$. 
We will associate with the Chern-Simons form the renormalized closed form $\bm{\frak{J}} = \frac{1}{\cN - 6} \frak{J}$.

It is now a straightforward task to construct ${\bm \S}_{\rm CS}$.
Using the constraints on the curvatures, we find for the first term in eq.
\eqref{eq:SigCS},
\be R^{\tilde{b}} \wedge \omega^{\tilde{a}} {\bm \G}_{\tilde{a} \tilde{b}} = - \RN^{IJ} \wedge \Phi_{IJ} + 2 \RK^a \wedge E_a - 4 \RS^{\a I} \wedge E_{\a I} \ . \label{eq:SigCS1}
\ee
Making use of the identities
\begin{subequations}
\begin{align} \RK^a \wedge E_a &= \ri E^a \wedge \frak{F}^{\a I} \wedge \frak{F}^\b_I (\g_a)_{\a\b} -  \ri E^\a_I \wedge E^{\b I} \wedge \frak{F}^a (\g_a)_{\a\b} - \rd (\frak{F}^a \wedge E_a) \ , \\
\RS^{\a I} \wedge E_{\a I} &= - \ri E^\a_I \wedge E^{\b I} \wedge \frak{F}^a (\g_a)_{\a\b} + \ri E^a \wedge \frak{F}^{\a I} \wedge \frak{F}^\b_I (\g_a)_{\a\b} + \rd (\frak{F}_\a^I \wedge E^\a_I) \ ,
\end{align}
\end{subequations}
which follow from eq. \eqref{eq:TRcsg} and the constraints \eqref{RPConstraint},
we can rewrite eq. \eqref{eq:SigCS1} as
\begin{align}
R^{\tilde{b}} \wedge \omega^{\tilde{a}} {\bm \G}_{\tilde{a} \tilde{b}} &=
	- R^{IJ} \wedge \Phi_{IJ} - 2 \ri E^a \wedge \frak{F}^{\a I} \wedge \frak{F}^\b_I (\g_a)_{\a\b}
	 \non\\&\quad
	+ 2 \ri E^\a_I \wedge E^{\b I} \wedge \frak{F}^a (\g_a)_{\a\b}+ {\rm exact \ form} \ .
\end{align}
The second term in eq. \eqref{eq:SigCS} is given by the sum of the following terms:
{\allowdisplaybreaks
\begin{subequations}
\begin{align}
\frac{1}{24} \Omega^{ab} \wedge \omega^{\tilde{b}} \wedge \omega^{\tilde{a}} (f_{\tilde{a} \tilde{b}}{}^{M_{cd}} {\bm \G}_{M_{cd} M_{ab}}) &= 
	- \frac{1}{6} \Omega^c \wedge \Omega^b \wedge \Omega^a \eps_{abc}
	+ \frac{2}{3} E^c \wedge \frak{F}^b \wedge \Omega^a \eps_{abc}
	\non\\&\quad
	+ \frac{2}{3} E^\a_I \wedge \frak{F}^{\b I} \wedge \Omega^a (\g_a)_{\a\b} \ ,
\, \\
\frac{1}{6} E^a \wedge \omega^{\tilde{b}} \wedge \omega^{\tilde{a}} (f_{\tilde{a} \tilde{b}}{}^{K_b} {\bm \G}_{K_b P_a}) &= 
	\frac{2}{3} E^c \wedge \frak{F}^b \wedge \Omega^a \eps_{abc}
	- \frac{2 \ri}{3} E^a \wedge \frak{F}^\a_I \wedge \frak{F}^{\b I} (\g_a)_{\a\b} 
	\non\\&\quad
	+ \frac{2}{3} E^a \wedge \frak{F}_a \wedge B
\ , \\
\frac{1}{6} \frak{F}^a \wedge \omega^{\tilde{b}} \wedge \omega^{\tilde{a}} (f_{\tilde{a} \tilde{b}}{}^{P_b} {\bm \G}_{P_b K_a}) &= 
	- \frac{2 \ri}{3} E^\a_I \wedge E^{\b I} \wedge \frak{F}^a (\g_a)_{\a\b}
	+ \frac{2}{3} E^c \wedge \frak{F}^b \wedge \Omega^a \eps_{abc}
	\non\\ &\quad
	+ \frac{2}{3} E^a \wedge \frak{F}_a \wedge B
\ , \\
\frac{1}{6} E^\a_I \wedge \omega^{\tilde{b}} \wedge \omega^{\tilde{a}} (f_{\tilde{a} \tilde{b}}{}^{S_\b^J} {\bm \G}_{S_\b^J Q_\a^I}) &=
	\frac{2}{3} E^\a_I \wedge \frak{F}^{\b I} \wedge \Omega^a (\g_a)_{\a\b}
	- \frac{4 \ri}{3} E^\a_I \wedge E^{\b I} \wedge \frak{F}^a (\g_a)_{\a\b}
	\non\\&\quad
	- \frac{4}{3} E^\a_I \wedge \frak{F}_{\a J} \wedge \Phi^{IJ}
	- \frac{2}{3} E^{\alpha}_I \wedge \frak{F}_\alpha^I \wedge B
\ , \\
\frac{1}{6} \frak{F}^\b_J \wedge \omega^{\tilde{b}} \wedge \omega^{\tilde{a}} (f_{\tilde{a} \tilde{b}}{}^{Q_\a^I} {\bm \G}_{Q_\a^I S_\b^J})  &= 
	- \frac{4 \ri}{3} E^a \wedge \frak{F}^\a_I \wedge \frak{F}^{\b I} (\g_a)_{\a\b}
	+ \frac{2}{3} E^\a_I \wedge \frak{F}^{\b I} \wedge \Omega^a (\g_a)_{\a\b}
	\non\\&\quad
	- \frac{4}{3} E^\a_I \wedge \frak{F}_{\a J} \wedge \Phi^{IJ}
	- \frac{2}{3} E^{\alpha}_I \wedge \frak{F}_\alpha^I \wedge B
\ , \\
\frac{1}{24} \Phi^{IJ} \wedge \omega^{\tilde{b}} \wedge \omega^{\tilde{a}} (f_{\tilde{a} \tilde{b}}{}^{N_{KL}} {\bm \G}_{N_{KL} N_{IJ}}) &= 
	 \frac{1}{3} \Phi^{IJ} \wedge \Phi_I{}^K \wedge \Phi_{KJ}
	- \frac{4}{3} E^\a_I \wedge \frak{F}_{\a J} \wedge \Phi^{IJ}
\ , \\
\frac{1}{6} B \wedge \omega^{\tilde{b}} \wedge \omega^{\tilde{a}} (f_{\tilde{a} \tilde{b}}{}^{\mathbb D} {\bm \G}_{\mathbb D \mathbb D})  &=
	\frac{2}{3} E^a \wedge {\frak{F}}_a \wedge B
	- \frac{2}{3} E^{\alpha}_I \wedge {\frak{F}}_\alpha^I \wedge B \ .
\end{align}
\end{subequations}}

The full Chern-Simons form is
\begin{align} {\bm \S}_{\rm CS} =& - \frac{1}{6} \Omega^c \wedge \Omega^b \wedge \Omega^a \eps_{abc} + 2 E^c \wedge \frak{F}^b \wedge \Omega^a \eps_{abc} 
- 4 \ri E^a \wedge \frak{F}^{\a I} \wedge \frak{F}^\b_I (\g_a)_{\a\b} \non\\
&+ 2 E^\a_I \wedge \frak{F}^{\b I} \wedge \Omega^a (\g_a)_{\a\b} - R^{IJ} \wedge \Phi_{IJ} + \frac{1}{3} \Phi^{IJ} \wedge \Phi_I{}^K \wedge \Phi_{KJ} \non\\
&- 4 E^\a_I \wedge \frak{F}_{\a J} \wedge \Phi^{IJ}
	+ 2 E^a \wedge \frak{F}_a \wedge B
	- 2 E^{\alpha}_I \wedge \frak{F}_\alpha^I \wedge B
	+ {\rm exact \ form} \ .
\end{align}
This result can be further simplified by using the explicit expressions for the curvatures
$\RM^{ab}$ and $\RN^{IJ}$ in eq. \eqref{eq:TRcsg}.
Since the superconformal Lorentz curvature $\RM^{ab}$ vanishes, we find
\begin{align} {\bm \S}_{\rm CS} &=
	- {\hat{R}}^a \wedge \Omega_a  -  \frac{1}{6} \Omega^c \wedge \Omega^b \wedge \Omega^a \eps_{abc}
	- 4 \ri E^a \wedge \frak{F}^{\a I} \wedge \frak{F}^\b_I (\g_a)_{\a\b}
	- {\hat{R}}^{IJ} \wedge \Phi_{IJ}
	\non\\ & \quad
	+ \frac{1}{3} \Phi^{IJ} \wedge \Phi_I{}^K \wedge \Phi_{KJ}
	+ 2 E^a \wedge \frak{F}_a \wedge B
	- 2 E^{\alpha}_I \wedge \frak{F}_\alpha^I \wedge B
	+ {\rm exact \ form} \ , \label{CSFORM}
\end{align}
where
\begin{align}
{\hat{R}}^{ab} := \rd \Omega^{ab} + \Omega^{ac} \wedge \Omega_c{}^b ~, \qquad
{\hat{R}}^{IJ} := \rd \Phi^{IJ} + \Phi^{IK} \wedge \Phi_{K}{}^J ~,
\end{align}
correspond to the Riemann and non-conformal $\rm SO(\cN)$ curvature tensors.

Some comments are necessary here about the simplifications which occur for
small values of $\cN$.
For $\cN = 1$, the $\rm SO(\cN)$ connection vanishes, giving
\begin{align} {\bm \S}_{\rm CS} &=
	- {\hat{R}}^a \wedge \Omega_a  -  \frac{1}{6} \Omega^c \wedge \Omega^b \wedge \Omega^a \eps_{abc}
	- 4 \ri E^a \wedge \frak{F}^{\a} \wedge \frak{F}^\b (\g_a)_{\a\b}
	\non \\ & \quad
	+ 2 E^a \wedge \frak{F}_a \wedge B
	- 2 E^{\alpha} \wedge \frak{F}_\alpha \wedge B
	+ {\rm exact \ form} \ .
\end{align}
Moreover, $\langle R^2 \rangle = 0$, and so the Chern-Simons form is closed,
\be \bm {\frak{J}} = {\bm \S}_{\rm CS}~,
\ee
without the need for the additional curvature induced form.
The corresponding action may be constructed straightforwardly.

The closed form $\bm{\frak J}$ was first constructed for $\cN=1$ in \cite{KT-M12} 
using the superspace formulation of \cite{KLT-M11}.
To compare that result to eq. \eqref{CSFORM}, we must degauge conformal
superspace to the superspace of \cite{KLT-M11}, following the procedure
detailed in \cite{BKNT-M1}. After adopting the gauge $B_A=0$,
all the special conformal connections may be expressed in terms of
additional torsion superfields, such as a real superfield $S$.
Adding to $\bm{\frak{J}}$ the exact form
\begin{align} 2 \rd (S E^a \wedge \Omega_a) =& \ 2 \rd S \wedge E^a \wedge \Omega_a + 2 S T^a \wedge \Omega_a - 2 S E^c \wedge \Omega^b \wedge \Omega^a \eps_{abc} \non\\
& + 2 S E^a \wedge E^\a \wedge \frak{F}^{\b} (\g_a)_{\a\b} \ ,
\end{align}
where $T^a$ is the torsion two-form of \cite{KLT-M11}, we find
\begin{align} \bm{\frak{J}} =& - \frac{1}{6} \Omega^c \wedge \Omega^b \wedge \Omega^a \eps_{abc} - 2 S E^c \wedge \Omega^b \wedge \Omega^a \eps_{abc} + 2 E^c \wedge \frak{F}^b \wedge \Omega^a \eps_{abc} \non\\
&+ 2 \rd S \wedge E^a \wedge \Omega_a -4 \ri E^a \wedge \frak{F}^\a \wedge \frak{F}^\b (\g_a)_{\a\b} + 2 E^\a \wedge \frak{F}^\b \wedge \Omega^a (\g_a)_{\a\b} \non\\
& + 2 S T^a \wedge \Omega_a + 2 S E^a \wedge E^\a_I \wedge \frak{F}^{\b I} (\g_a)_{\a\b} + {\rm exact \ form} \ .
\end{align}
This result can be shown to match the closed form given in \cite{KT-M12} up to 
an exact form (and up to conventions).

There are also some simplifications which occur for $\cN = 2$.
The term with three $\rm SO(2)$ connections vanishes identically, giving
\begin{align}
{\bm \S}_{\rm CS} &= 
	- {\hat{R}}^a \wedge \Omega_a  -  \frac{1}{6} \Omega^c \wedge \Omega^b \wedge \Omega^a \eps_{abc}
	- 4 \ri E^a \wedge \frak{F}^{\a I} \wedge \frak{F}^\b_I (\g_a)_{\a\b} \non\\
	&\quad
	- {{\hat{R}}}^{IJ} \wedge \Phi_{IJ}
	+ 2 E^a \wedge \frak{F}_a \wedge B
	- 2 E^{\alpha}_I \wedge \frak{F}_\alpha^I \wedge B
	+ {\rm exact \ form} \ .
\end{align}
We again find $\langle R^2 \rangle = 0$ since the $\rm SO(2)$ curvature only appears
at dimension $2$ in the covariant derivative algebra \eqref{N=2Algebra}. 
The Chern-Simons form is again closed without the need to introduce a curvature induced form.

For $\cN > 2$ one finds (using the constraints on the curvatures as well
as on the Cartan-Killing metric) that
\be \langle R^2 \rangle = (6 - \cN) \RN^{IJ} \wedge \RN_{IJ}~,
\ee
which is non-vanishing in general.
Hence, it is necessary to introduce the curvature induced three-form.
In the next subsection, we will explicitly construct such a three-form
for $\cN=3$, $4$, and $5$.

%%%%%%%%%%%%%%%%%%%%%%%%%%%%%%%%%%%%%%%%%%%%%%%%%%%%%%

\subsection{The curvature induced three-form}

We wish to find a solution to\footnote{In this subsection all wedge products are implicit.}
\be \rd \S_{R} = (6 - \cN) \RN^{IJ} \RN_{IJ} \ ,
\ee
where $\S_R$ is a covariant three-form built entirely out of the curvature components.
For our construction we will find it is useful to use the renormalized curvature 
induced form
\be \bm{\S}_{R} = \frac{1}{\cN - 6} \S_R \ ,
\ee
which satisfies the superform equation
\bea
\rd {\bm{\S}}_R = - \RN^{IJ} \RN_{IJ} \ . \label{SigmaRren}
\eea
Then the closed form $\bm{\frak{J}}$ is given by
\be \bm{\frak{J}} = {\bm \S}_{\rm CS} - {\bm \S}_R \ .
\ee

%%%%%%%%%%%%%%%%%%%%%%%%%%%%%%%%%%%%%%%%%%%%%%%%%%%%%%

\subsubsection{The $\cN = 3$ case}

The expression on the right hand side of eq. \eqref{SigmaRren} is
\begin{align}
\RN^{IJ} \RN_{IJ}=&~
\frac{1}{4} E^\d_LE^\g_KE^bE^a\eps_{abc}\Big{[}
16 \d^{KL}(\g^c)_{\g\d}W^\r W_\r
\Big{]}
\non\\
&~
+\frac{1}{6}E^\d_L E^c E^b E^a\ve_{abc}
 \Big{[}
- 8\ri  \nabla_\d^{L}(W^\g W_\g)
 \Big{]}
~,
\end{align}
which involves only the $\cN=3$ super Cotton tensor $W_\alpha$.
The curvature induced three-form ${\bm \S}_R$ should be a solution 
to eq. \eqref{SigmaRren},
\bea\label{eq:SigmaRrenComp}
- (\RN^{KL} \RN_{KL})_{ABCD}=
4\nabla_{[A}{\bm \S}_{BCD\}}
+ 6T_{[AB}{}^E{\bm \S}_{|E|CD\}}
~.
\eea
where ${\bm\S}_{ABC} := (\bm \S_R)_{ABC}$ involves only the components
of $W_\alpha$ itself and transforms covariantly under $\cH$.
Taking into account the dimension of $W_\alpha$,
the only possible solution is
\bea \label{sigmaN=3}
{\bm \S}_{\a}^I{}_{\b}^J{}_{\g}^K
&=&
{\bm \S}_a{}_{\b}^J{}_{\g}^K
={\bm \S}_{ab}{}_{\g}^K
=0~, \qquad
{\bm \S}_{abc} = - 8\,\ri\,\ve_{abc}W^\g W_\g \ ,
\eea
where the constant of proportionality is set by explicitly checking eq. \eqref{eq:SigmaRrenComp}. 
It is easy to see by inspection
that the three-form ${\bm \S}_R$ is indeed $\cH$-invariant.

%%%%%%%%%%%%%%%%%%%%%%%%%%%%%%%%%%%%%%%%%%%%%%%%%%%%%%

\subsubsection{The $\cN > 3$ case}

For all cases $\cN > 3$, the superspace geometry involves the super
Cotton tensor $W^{IJKL}$, and the expression on the right hand side of eq. \eqref{SigmaRren} is
generically given by
{\allowdisplaybreaks
\begin{align}
\RN^{IJ} &\RN_{IJ}
=~
\frac{1}{24}E^\d_L E^\g_K E^\b_J E^\a_I\Big{[}
-24\eps_{\a\b}\eps_{\g\d}W^{PQIJ}W^{KL}{}_{PQ}
\Big{]}
\non\\
&~
+\frac{1}{6}E^\d_L E^\g_K E^\b_JE^a\Big{[}
\frac{12\ri}{(\cN-3)}\eps_{\g\d}(\g_a)_{\b\r}W^{PQKL}(\nabla_I^\r W_{PQ}{}^{JI})
\Big{]}
\non\\
&~
+\frac{1}{4} E^\d_L E^\g_KE^b E^a\eps_{abc}\Big{[}\,
-\frac{2}{(\cN-3)^2}\eps_{\g\d}(\g^c)_{\r\t}(\nabla_I^\r W^{PQKI})(\nabla_J^\t W_{PQ}{}^{LJ})
\non\\
&~~~~~~~~~~~~~~~~~~~~~~~~~
-\frac{2}{(\cN-3)^2}(\g^c)_{\g\d}(\nabla_I^\r W^{PQKI})(\nabla_{\r J} W_{PQ}{}^{LJ})
\non\\
&~~~~~~~~~~~~~~~~~~~~~~~~~
-\frac{2}{(\cN-2)(\cN-3)}\eps_{\g\d}(\g^c)_{\r\t}W_{PQ}{}^{KL}(\nabla^\r_I\nabla^\t_J W^{PQIJ})
\Big{]}
\non\\
&~
+\frac{1}{6}E^\d_L E^c E^bE^a\eps_{abc}\Big{[}
-\frac{2\ri}{(\cN-2)(\cN-3)^2}(\nabla_I^\r W^{PQLI})(\nabla_{\d M}\nabla_{\r N} W_{PQ}{}^{MN})
\Big{]}
~. \label{generalR2}
\end{align}}
Now it is not so straightforward to solve eq. \eqref{SigmaRren},
\bea
- (\RN^{KL} \RN_{KL})_{ABCD}=
4\nabla_{[A}{\bm \S}_{BCD\}}
+ 6T_{[AB}{}^E{\bm \S}_{|E|CD\}} \ . \label{RinducedI}
\eea
This is because if we require ${\bm \S}_R$ to be built only out of $W^{IJKL}$, then
the only possible ansatz for the lowest components is
\be {\bm \S}_\a^I{}_\b^J{}_\g^K = 0 \ , \quad {\bm \S}_a{}_\b^J{}_\g^K = \ri \,(\g_a)_{\b\g}\Big(
A\, \d^{JK} W^{ILPQ} W_{ILPQ}
+ B\, W^{LPQJ} W_{LPQ}{}^K\Big)
 \ ,
 \label{ansatz}
\ee
where $A$ and $B$ are unknown constants.\footnote{The ansatz is quadratic in the super Cotton tensor
and such that ${\bm \S}_\a^I{}_\b^J{}_\g^K = 0$. One might wonder whether contributions linear in
$W^{IJKL}$ can occur. In this case, the lowest component of ${\bm \S_R}$ should be non-zero and
proportional to $\nabla_{\a P}W^{JKLP}$. However,
since the only possible algebraic combination is
${\bm \S}_{\a}^I{}_\b^J{}_{\g}^K\propto\ve_{[\a\b}\nabla_{\g] P}W^{JKLP}=0$,
this possibility is ruled out.}
By using eqs. \eqref{ansatz} and \eqref{generalR2},
the lowest dimensional equation in eq. \eqref{RinducedI} turns out to be equivalent to the following equation
\bea
0&=&E^\d_LE^\g_KE^\b_JE^\a_I\,\ve_{\a\b}\ve_{\g\d}\Big(
- W^{PQIJ}W^{KL}{}_{PQ}
+A\,W^{PQRS}W_{PQRS}\d^{J[K}\d^{L]I}
\non\\
&&~~~~~~~~~~~~~~~~~~~~~~~~~~
+B\,W^{PQRJ}W_{PQR}{}^{[K}\d^{L]I}
\Big) \label{4.28}
~.
\eea
For $\cN > 5$, the first term in this equation contains a double traceless contribution of the form
\bea
\Big(\d^R_{[K}\d^{[I}_{|S|}-\frac{1}{\cN}\d^R_{S}\d^{[I}_{[K}\Big)
\Big(\d^{|T|}_{L]}\d^{J]}_{U}-\frac{1}{\cN}\d^{J]}_{L]}\d^T_{U}\Big)
W^{SUPQ}W_{RTPQ}
~,
\label{WHAT-IS-IT?}
\eea
which cannot be cancelled by the second and third terms in eq. \eqref{4.28}. As a result, 
we specialize to the $\cN = 4$ and $\cN = 5$ cases here where this contribution is identically zero.

%%%%%%%%%%%%%%%%%%%%%%%%%%%%%%%%%%%%%%%%%%%%%%%%%%%%%%

\subsubsection{The $\cN = 4$ case}

In this case, the super Cotton tensor can be written as
\be W^{IJKL} = \eps^{IJKL} W \ .
\ee
Then eq. \eqref{generalR2} reduces to
\begin{align}
&\RN^{IJ} \RN_{IJ}
=~
\frac{1}{24}E^\d_L E^\g_K E^\b_J E^\a_I\Big{[}
-96\eps_{\a\b}\ve_{\g\d}\d^{I[K}\d^{L]J}W^2
\Big{]}
\non\\
&~
+\frac{1}{6}E^\d_L E^\g_K E^\b_JE^a\Big{[}
48\ri\ve_{\g\d}(\g_a)_{\b\r}\d^{J[K}(\nabla^{\r L]} W)W
\Big{]}
\non\\
&~
+\frac{1}{4} E^\d_L E^\g_KE^b E^a\eps_{abc}\Big{[}\,
-4\ve_{\g\d}(\g^c)^{\r\t}(\nabla_\r^{K} W)(\nabla_\t^{L} W)
-4(\g^c)_{\g\d}\d^{KL}(\nabla_P^\r W)(\nabla_{\r}^{P} W)
\non\\
&\,~~~~~~~~~~~~~~~~~~~~~~~~~~~~
+4(\g^c)_{\g\d}(\nabla^{\r K} W)(\nabla_{\r}^{L} W)
-4\eps_{\g\d}(\g^c)^{\r\t}(\nabla_\r^{[K}\nabla_\t^{L]} W)W
\Big{]}
\non\\
&~
+\frac{1}{6}E^\d_L E^c E^bE^a\eps_{abc}\Big{[}
-4\ri(\nabla_P^\r W)(\nabla_{\d}^{[L}\nabla_{\r}^{P]} W)
\Big{]}
~.
\end{align}
The ansatz
\begin{align}
{\bm \S}_\a^I{}_\b^J{}_\g^K = 0 \ , \qquad
{\bm \S}_a{}_\b^J{}_\g^K  = B\,\ri \,\d^{JK} (\gamma_a)_{\b\g} \,W^2 \ ,
\end{align}
with an undetermined constant $B$ turns out to solve the constraint \eqref{RinducedI}. Using the following consequences of eq. \eqref{2.39}
\bsubeq
\begin{align} \nabla_\a^I\nabla_{\b}^JW =&
\nabla_{(\a}^{[I}\nabla_{\b)}^{J]}W
+\ri\d^{IJ}\nabla_{\a\b} W
+\frac{1}{8}\eps_{\a\b}\d^{IJ}\nabla^{\g}_{P}\nabla_{\g}^{P}W \ , \\
\nabla_\a^I \nabla_{\b}^{[J}\nabla_{\g}^{K]}W
=&~
-\frac{1}{6}\ve^{IJK}{}_{M}\eps^{MLPQ}\nabla_{(\a L}\nabla_{\b P}\nabla_{\g) Q}W
\non\\
&~
+2\ri\d^{I[J}\nabla_{(\a\b}\nabla_{\g)}^{K]} W
-\frac{8}{3}\ri\d^{I[J}\eps_{\a(\b}\nabla_{\g)\d}\nabla^{\d K]} W
~, \\
\nabla_\a^{I}\nabla^{\r}_{Q}\nabla_{\r}^{Q}W
=& -8\ri\nabla_{\a\d}\nabla^{\d I} W
~,
\end{align}
\esubeq
we find the solution
\begin{align}  \label{sigmaN=4}
{\bm \S}_\a^I{}_\b^J{}_{\g}^K &= 0
~,\qquad
{\bm \S}_a{}_\b^J{}_{\g}^K=~
- 4\ri\,(\g_a)_{\b\g}\d^{JK}W^2
~,\qquad
{\bm \S}_{ab}{}_{\g}^K
= 2\ve_{abc}(\g^c)_{\g\d}(\nabla^{\d K}W^2)
~,
\non\\
{\bm \S}_{abc}
&= 2\ri\,\ve_{abc}(\nabla^\r_P W)(\nabla_\r^P W)
+\frac{\ri}{2}\ve_{abc}W(\nabla^\r_P\nabla_\r^P W)
~.
\end{align}

%%%%%%%%%%%%%%%%%%%%%%%%%%%%%%%%%%%%%%%%%%%%%%%%%%%%%%

\subsubsection{The $\cN = 5$ case}

The super Cotton tensor may now be written as
\be W^{IJKL} = \eps^{IJKLP} W_P \ ,
\ee
and eq. \eqref{generalR2} becomes
\begin{align}
\RN^{IJ} &\RN_{IJ}
=~
\frac{1}{24}E^\d_L E^\g_K E^\b_J E^\a_I\Big{[}
-96\ve_{\a\b}\ve_{\g\d}\Big(
\d^{IK}\d^{JL}W^2
-2\d^{IK}W^JW^L
\Big)
\Big{]}
\non\\
&~
+\frac{1}{6}E^\d_L E^\g_K E^\b_JE^a\Big{[}
24\ri\ve_{\g\d}(\g_a)_{\b\r}\Big(
2\d^{JK}(\de^{\r[L} W^{S]})W_S
+(\de^{\r[K} W^{L]})W^J
\Big)
\Big{]}
\non\\
&~
+\frac{1}{4} E^\d_L E^\g_KE^b E^a\ve_{abc}\Big{[}\,
-4\ve_{\g\d}(\g^c)_{\r\t}\d^{KP}(\de_{[P}^\r W_{Q]})(\de^{\t[L} W^{Q]})
\non\\
&~~~~~~~~~~~~~~~~~~~~~~~~~~~~
-2(\g^c)_{\g\d}\d^{KL}(\de_{[P}^\r W_{Q]})(\de_{\r}^{[P} W^{Q]})
\non\\
&~~~~~~~~~~~~~~~~~~~~~~~~~~~~
+4(\g^c)_{\g\d}\d^{KP}(\de_{[P}^\r W_{Q]})(\de_{\r}^{[L} W^{Q]})
\non\\
&~~~~~~~~~~~~~~~~~~~~~~~~~~~~
-4\ve_{\g\d}(\g^c)^{\r\t}(\de_\r^{[K}\de_\t^{L} W^{P]})W_P
\Big{]}
\non\\
&~
+\frac{1}{6}E^\d_L E^c E^bE^a\ve_{abc}\Big{[}
-2\ri(\de_{[P}^\r W_{Q]})(\de_{\d}^{[L}\de_{\r}^{P} W^{Q]})
\Big{]}
~.
\end{align}

We make the ansatz
\bea
{\bm \S}_\a^I{}_\b^J{}_{\g}^K=0
~,
\quad
{\bm \S}_a{}_\b^J{}_{\g}^K=\ri(\g_a)_{\b\g}\Big(B\d^{JK}W^2+CW^{J}W^K\Big) \ , \quad W^2 := W^I W_I
\eea
with $B$ and $C$ arbitrary constants. With the help of the identities
\bsubeq
\begin{align} \de_\a^IW^J &=
\frac{1}{5}\d^{IJ}\de_\a^{P}W_{P}
+\de_{\a}^{[I}W^{J]} \ , \\
\de^{\g J}\de_{\g [J}W_{I]}
&=
\frac{4}{3}\de^\g_{ P}\de_{\g}^{P}W_I \ , \\
\de_\a^I\de_{\b}^{ [J}W^{K]}
&=
\de_{(\a}^{ [I}\de_{\b)}^{J}W^{K]}
-\frac{1}{3}\ve_{\a\b}\de^\g_{ P}\de_{\g}^{P}W^{[J}\d^{K]I}
-2\ri\de_{\a\b} W^{[J}\d^{K]I}
~, \\
\de_\a^I\de_\b^{J}W_J
&=
-\frac{5}{6}\ve_{\a\b}\de^{\g}_P\de_{\g}^{P}W^I
+5\ri\de_{\a\b}W^I \ , \\
\de_\a^{I}\de^{\g}_P\de_{\g}^PW^J
&=
6\ri\de_{\a}{}^{\b}\de_{\b}^{[I}W^{J]}
-\frac{6}{5}\ri\d^{IJ}\de_{\a}{}^{\b}\de_\b^{P}W_P
+3\ri \ve^{IJKLP}W_K \de_{\a L}W_{P} \ ,
\end{align}
\esubeq
which are consequences of eq. \eqref{CSBIN>3} for $\cN = 5$, we find the solution
\bsubeq  \label{sigmaN=5}
\begin{align}
{\bm \S}_\a^I{}_\b^J{}_{\g}^K
=&~0
~,
\\
{\bm \S}_a{}_\b^J{}_{\g}^K
=&~
- \ri(\g_a)_{\b\g}\Big(4\d^{JK}W^2-8W^{J}W^K\Big)
~,
\\
{\bm \S}_{ab}{}_{\d}^L
=&~
- 4\ve_{abc}(\g^c)_{\d}{}^{\r}\Big(
(\de_\r^{[L} W^{S]})W_S
-\frac{1}{5}(\de_\r^{P}W_P)W^L
\Big)
~,
\\
{\bm \S}_{abc}
=&~
- \ri\ve_{abc}\Big(
\frac{2}{25}(\de^{\r P}W_P)(\de_\r^QW_Q)
-(\de_{[P}^\r W_{Q]})(\de_{\r}^{[P} W^{Q]})
-\frac{2}{3}(\de^\r_{ P}\de_{\r}^{P}W^{S})W_S
\Big) \ .
\end{align}
\esubeq

With the closed three-forms constructed in this section, we can 
build the off-shell $\cN<6$ conformal supergravity actions. So far, the only
missing ingredient to the construction is the explicit component structure
of the $\cN<6$ Weyl multiplets.
This will be our goal in the next section.

%%%%%%%%%%%%%%%%%%%%%%%%%%%%%%%%%%%%%%%%%%%%%%%%%%%%%%
%%%%%%%%%%%%%%%%%%%%%%%%%%%%%%%%%%%%%%%%%%%%%%%%%%%%%%

\section{The Weyl multiplet} \label{compAnalysis}

In this section, we elaborate on the component structure of the
conformal superspace of \cite{BKNT-M1}, which will correspond
to the Weyl multiplet. Although we will
mainly be interested in the cases where $\cN<6$, for which
we can explicitly construct the off-shell conformal supergravity action,
the results here will hold for general $\cN$.

%%%%%%%%%%%%%%%%%%%%%%%%%%%%%%%%%%%%%%%%%%%%%%%%%%%%%%

\subsection{Component fields}

The $\cN$-extended Weyl multiplet in three dimensions may be extracted from the superspace structure via component projections. 
It involves a set of gauge one-forms: the vielbein $e_m{}^a$, the gravitino $\psi_m{}^\a_I$, the $\rm SO(\cN)$ gauge field 
$V_m{}^{IJ}$ and the dilatation gauge field $b_m$. They appear in the superspace formulation as the lowest components of their corresponding superforms,
\begin{align}\label{eq:GaugeFields}
e_m{}^a &:= E_m{}^a| \ , \qquad \psi_m{}^\a_I := 2 E_m{}^\a_I| ~, \qquad
V_m{}^{IJ} := \Phi_m{}^{IJ}| \ ,\qquad b_m := B_m| ~,
\end{align}
where 
the bar 
projection \cite{WZ2, WB, GGRS}
of a superfield $V(z) =V(x,\q)$ is defined by 
the standard rule $V| := V(x,\q)|_{\theta = 0}$.
The remaining connection fields turn out to be \emph{composite}, and are built out of other fields.
These are the spin connection  $\omega_m{}^{ab}$
and the special conformal and $S$-supersymmetry connections $\frak{f}_m{}^a$ and $\phi_m{}_\a^I$,
\begin{align}\label{eq:compGaugeFields}
\omega_m{}^{ab} := \Omega_m{}^{ab}| \ , \quad  
\frak{f}_m{}^a := \frak{F}_m{}^a| \ , \quad
\phi_m{}^I_\a := 2 \frak F_m{}^I_\a| \ .
\end{align}

Using the higher-$\theta$ parts of the superspace diffeomorphisms and
$\cH$-gauge transformations, we can impose a Wess-Zumino gauge where we
fix the $\theta$ expansions of the super one-forms, as well as the
lowest components of the spinor vielbein  $E_\mu{}^A$ and
connection $\omega_\mu{}^{\ul a}$,
so that they are completely determined by the lowest components of the
fields defined in eqs. \eqref{eq:GaugeFields} and \eqref{eq:compGaugeFields}
as well as the components of the super
Cotton tensor.\footnote{Technically this is equivalent to making use of the so-called double bar projection
\cite{Baulieu:1986dp,Binetruy:2000zx}.}
Therefore, this is the entire physical field
content of the superspace geometry of \cite{BKNT-M1}.

For $\cN = 1$ and $\cN = 2$ it is possible to show that the number of
bosonic and fermionic degrees of freedom in eq. \eqref{eq:GaugeFields}
are the same without the need to introduce additional
fields \cite{vanN85, RvanN86}. However, for $\cN > 2$ such fields are necessary
to ensure the theory is off-shell.  Since one can deduce the lower $\cN$ cases
from the $\cN > 3$ case, we will first focus on the $\cN > 3$ case.

For $\cN > 3$ the additional fields are encoded in the super Cotton tensor $W^{IJKL}$ \cite{GreitzHowe,GGHN}.
The independent fields may be deduced by taking spinor derivatives of 
$W^{IJKL}$ and using the Bianchi identity \eqref{CSBIN>3},
to eliminate algebraically dependent combinations.
We then define the component fields as\footnote{It is possible to show
that one can omit the symmetrization in the definition of $w_{\a\b}{}^{IJ}$, $w_{\a\b\g}{}^I$ and 
$w_{\a\b\g\d}$.}
{\allowdisplaybreaks
\begin{subequations} \label{defCompW}
\begin{align}
w_{IJKL} &:= W_{IJKL}| \ , \\
w_\a{}^{IJK} &:= - \frac{\ri}{2 (\cN - 3)} \nabla_{\a L} W^{IJKL}| \ , \\
w_{\a\b}{}^{IJ} &:= \frac{\ri}{2 (\cN - 2) (\cN - 3)} \nabla_{(\a K} \nabla_{\b) L} W^{IJKL}| \ , \\
w_{\a\b\g}{}^I &:= \frac{\ri}{(\cN - 1) (\cN -2) (\cN - 3)} \nabla_{(\a J} \nabla_{\b K} \nabla_{\g ) L} W^{IJKL}| \ , \\
w_{\a\b\g\d} &:= - \frac{1}{\cN (\cN - 1) (\cN -2) (\cN - 3)}  \nabla_{(\a I} \nabla_{\b J} \nabla_{\g K} \nabla_{\d ) L} W^{IJKL}| \ , \\
y^{I J K L} &:= \frac{\ri}{\cN - 3} \nabla^{\g [I}  \nabla_{\g P} W^{JKL]P}| \ , \\
X_{\a_1 \cdots \a_n}{}^{I_1 \cdots I_{n+4}} &:= I(n) \nabla_{(\a_1}^{[I_1} \cdots \nabla_{\a_n)}^{I_n} W^{I_{n+1} \cdots I_{n+4}]}| \ , \label{eq:defHigherX}
\end{align}
\end{subequations}}
where we define the factor $I(n)$ by\footnote{The factor $I(n)$ is needed to ensure the fields $X_{\a_1 \cdots \a_n}{}^{I_1 \cdots I_{n+4}}$ are real.}
\be
I(n)=
\begin{cases}
\ri \ , & n = 1,2 \ ({\rm mod} \ 4) \\
1 \ , & n = 3,4 \ ({\rm mod} \ 4) ~.
\end{cases}
\ee
These fields, when organized by dimension, diagrammatically 
form the following tower \cite{GreitzHowe,GGHN}:\\
\begin{minipage}[t]{\textwidth}
\begin{picture}(430,195)
\put(197,170){$w^{I_1\cdots I_4}$}
\put(205,165){\vector(-1,-1){20}}
\put(210,165){\vector(1,-1){20}}
\put(165,130){$X_\a{}^{I_1\cdots I_5}$}
\put(230,130){$w_\a{}^{I_1I_2I_3}$}
\put(165,125){\vector(-1,-1){20}}
\put(175,125){\vector(1,-1){20}}
\put(240,125){\vector(-1,-1){20}}
\put(250,125){\vector(1,-1){20}}
\put(125,90){$X_{\a_1\a_2}{}^{I_1\cdots I_6}$}
\put(197,90){$y^{I_1\cdots I_4}$}
\put(270,90){$w_{\a_1\a_2}{}^{I_1I_2}$}
\put(122,80){\vector(-1,-1){20}}
\put(292,80){\vector(1,-1){20}}
\put(83,45){$\cdots$
}
\put(310,45){$w_{\a_1\a_2\a_3}{}^{I_1}$}
\put(80,38){\vector(-1,-1){20}}
\put(335,35){\vector(1,-1){20}}
\put(40,0){$X_{\a_1\cdots\a_{\cN-4}}{}^{I_1\cdots I_{\cN}}$
}
\put(350,0){$w_{\a_1\cdots\a_4}$}
\end{picture}
\begin{center}{\bf Figure 1.} $\cN$-extended super Cotton tensor\end{center}
\end{minipage}
The arrows connecting the various independent components correspond to the action of $Q$-supersymmetry, which 
raises the dimension by $1/2$ as we proceed downward.
The left hand branch terminates at $n = \cN - 4$, and therefore the first term of this branch
shows up at $\cN = 5$. 
The right hand branch, however, is composed of the projections of curvatures that appear in the covariant derivative algebra. As a result, 
the components fields $w_{\a\b}{}^{IJ}$, $w_{\a\b\g}{}^I$ and $w_{\a\b\g\d}$ are constrained by the geometry to be composite. To see this, 
we note that they may be expressed in terms of the curvatures appearing in the commutator of two vector covariant derivatives,\footnote{The curvatures appearing in these expressions are
Hodge duals, using the normalization $F^a = \frac{1}{2} \eps^{abc} F_{bc}$ for a two-form
$F_{bc}$ \cite{BKNT-M1}.}
\begin{align} w_c{}^{IJ} &= - \hf \RN_c{}^{IJ}| \ , \quad w_c{}_\g^K = - 2 \RS_c{}_\g^K| \ , \non\\
 w_{ab} &:= \frac{1}{4} (\g_a)^{\a\b} (\g_b)^{\g\d} w_{\a\b\g\d} = - 2 \RK_{a , b}| \ .
\end{align}
Their component expressions are derived in Appendix \ref{compAnalysis2}.
The case $\cN=8$ is especially interesting since one can impose either a
self-dual or anti-self-dual condition on the super Cotton tensor; this amounts
to equating the terms in the left and right branches \cite{GGHN}.

It should be mentioned that the coefficients in the definitions \eqref{defCompW} 
were chosen so that  we can easily extract the component results for lower $\cN$ from 
the higher ones. All we must do is follow the prescription given in section 4 of Ref. \cite{BKNT-M1}. We independently switch off the components with more than 
$\cN$ $\rm SO(\cN)$ indices and define
\be \eps^{I_1 \cdots I_{\cN}} w_{\a_1 \cdots \a_{4 - \cN}} := w_{\a_1 \cdots \a_{4 - \cN}}{}^{I_1 \cdots I_{\cN}} .
\ee
Then we find that the $\cN = 1$ components of the Cotton tensor are 
\be w_{\a\b\g} := W_{\a\b\g}| \ , \quad w_{\a\b\g\d} := \ri \nabla_{(\a} W_{\b\g\d)}| \ ,
\ee
while for $\cN = 2$ we have
\be w_{\a\b} := W_{\a\b}| \ , \quad w_{\a\b\g}{}^I := 2 \eps^{IJ} \nabla_{(\a J} W_{\b\g)}| \ , \quad w_{\a\b\g\d} := \ri \eps_{IJ} \nabla_{(\a}^{I} \nabla_\b^{J} W_{\g\d )}| \ , 
\ee
which must all be composite. For $\cN = 3$ the component fields of the super Cotton tensor are
\begin{subequations}\label{eq:CompN=3}
\begin{gather}
w_\a := W_\a| \ , \quad
w_{\a\b}{}^{IJ} := - \eps^{IJK} \nabla_{(\a K} W_{\b)}| \ , \quad
w_{\a\b\g}{}^I := - \eps^{IJK} \nabla_{(\a J} \nabla_{\b K} W_{\g)}| \ , \\
w_{\a\b\g\d}{} := - \frac{\ri}{3} \eps_{IJK} \nabla_{(\a}^I \nabla_\b^J \nabla_\g^K W_{\d)}| \ ,
\end{gather}
\end{subequations}
where the only auxiliary field is $w_\a$ and all other components are composite.

Before moving on, we would like to mention that the supersymmetry transformations of the 
component fields may be derived efficiently from conformal superspace. In Appendix \ref{SUSY} we give the 
supersymmetry transformations relevant to our considerations.

%%%%%%%%%%%%%%%%%%%%%%%%%%%%%%%%%%%%%%%%%%%%%%%%%%%%%%

\subsection{Analysis of the curvature constraints}

We have already mentioned that in the covariant derivative algebra \eqref{covDN>3},
the torsion tensor takes its constant value, while the Lorentz and
dilatation curvatures vanish. These constraints imply certain relations
on the gauge fields: in particular, they algebraically constrain the spin connection
$\omega_m{}^{ab}$, the $S$-supersymmetry connection $\phi_m{}^\alpha_I$
and the special conformal connection ${\frak f}_m{}^a$ to be composite.
In this subsection, we analyze these curvature constraints and give the
explicit algebraic solutions for the composite connections in our
conventions.

%%%%%%%%%%%%%%%%%%%%%%%%%%%%%%%%%%%%%%%%%%%%%%%%%%%%%%

\subsubsection{Vector torsion}

The vector torsion is given by eq. \eqref{Tor1}.
The projection of its lowest component is
\be T_{mn}{}^c| = 2 \partial_{[m} e_{n]}{}^c + 2 \omega_{[mn]}{}^c + 2 b_{[m} e_{n]}{}^c \ .
\ee
Since the only non-vanishing covariant torsion is $T_\a^I{}_\b^J{}^c = -2 \ri \d^{IJ} (\g^c)_{\a\b}$, we find\footnote{We use here the identity $F_{mn} = E_m{}^A E_n{}^B F_{AB} (-1)^{\eps_A \eps_B}$
for a two-form $F$.}
\be T_{mn}{}^c| = -\frac{\ri}{2} (\psi_m^I \g^c \psi_n{}_I) \ ,
\ee
so we have
\be - \cC_{abc}+ 2 \omega_{[ab]c} + 2 b_{[a} \eta_{b] c} =  -\frac{\ri}{2} (\psi_a^I \g_c \psi_b{}_I) \ ,
\ee
where $\psi_a{}_\a^I := e_a{}^m \psi_m{}_\a^I$ and $\cC_{abc} := - 2 e_a{}^m e_b{}^n \partial_{[m} e_{n] c}$. This allows us to solve for the spin connection in terms of the vielbein, gravitino 
and dilatation connection,
\begin{align} \omega_{abc} =& \ \omega(e)_{abc} - \frac{\ri}{4} (\psi_a{}^I \g_c \psi_b{}_I - \psi_b{}^I \g_a \psi_{c I} + \psi_c{}^I \g_b \psi_{a I}) + 2 b_{[b} \eta_{c ] a} \ ,
\end{align}
where
\be \omega(e)_{abc} = \frac{1}{2} (\cC_{abc} + \cC_{cab} - \cC_{bca})
\ee
is the contribution to the spin connection solely from the vielbein.

%%%%%%%%%%%%%%%%%%%%%%%%%%%%%%%%%%%%%%%%%%%%%%%%%%%%%%

\subsubsection{Spinor torsion}

The spinor torsion is given by eq. \eqref{Tor2}, and its 
projection to lowest component is
\be T_{mn}{}^\a_I| = \cD_{[m} \psi_{n]}{}^\a_I + \ri \phi_{[m}{}^\b_I (\g_{n]})_\b{}^\a  \ ,
\ee
where
\be \cD_{m} \psi_{n}{}^\a_I = \partial_{m} \psi_{n}{}^c + \frac{1}{2} \psi_{n}{}^\b_I \omega_{m}{}^c (\g_c)_\b{}^\a + \hf b_m \psi_n{}^\a_I - V_m{}^{IJ} \psi_n{}^\a_J
\ee
and $\g_m = e_m{}^a \g_a$.

Now since the spinor torsion vanishes at all mass dimensions,
\be T_{mn}{}^\a_I| = 0 \ ,
\ee
one finds a relation between the gravitino field strength and the
$S$-supersymmetry connection:
\be \Psi_{ab}{}^\a_I := 2 e_{[a}{}^m e_{b]}{}^n \cD_m \psi_n{}^\a_I = - 2 \ri \phi_{[a}{}^\b_I (\g_{b]})_\b{}^\a \ , \quad \phi_a{}^\b_J := e_a{}^m \phi_m{}^\b_J \ .
\ee
This equation algebraically determines $\phi_a{}^\a_I$:
\be \phi_a{}^\a_I = \frac{\ri}{2} (\g^b)_\b{}^\a \Psi_{ab}{}^\b_I + \frac{\ri}{4} \eps_a{}^{bc} \Psi_{bc}{}^\a_I \ . \label{confPhi}
\ee

%%%%%%%%%%%%%%%%%%%%%%%%%%%%%%%%%%%%%%%%%%%%%%%%%%%%%%

\subsubsection{Dilatation curvature}

Taking the projection of the dilatation curvature \eqref{Dcurv}, we find
\be \RD_{mn}| = 2 \partial_{[m} b_{n]} + 4 \frak{f}_{[mn]} + \psi_{[m}{}^\a_I \phi_{n]}{}_\a^I \ .
\ee
Because this quantity is constrained to vanish,
\be \RD_{mn}| = 0 \ ,
\ee
the antisymmetric part of the special conformal connection is determined
to be
\be \frak{f}_{[ab]} = - \hf e_a{}^m e_b{}^n \partial_{[m} b_{n]} - \frac{1}{4} \psi_{[a}{}^\a_I \phi_{b]}{}_\a^I \ . \label{DilEqfF}
\ee

%%%%%%%%%%%%%%%%%%%%%%%%%%%%%%%%%%%%%%%%%%%%%%%%%%%%%%

\subsubsection{Lorentz curvature}

Finally, we address the Lorentz curvature, eq. \eqref{MCurv}. Its projection
can be written
\be \RM_{mn}{}^{ab}| = 2 \partial_{[m} \omega_{n]}{}^{ab} - 2 \omega_{[m}{}^{ac} \omega_{n]}{}_c{}^b + 8 e_{[m}{}^{[a} \frak{f}_{n]}{}^{b]} + \psi_{[m}{}^\a_I \phi_{n]}{}^{\b I} (\g_c)_{\a\b} \eps^{cab}~. \\
\ee
Constraining this to vanish leads to
\be 0 = \RM_{ab}{}^{cd}| = \cR_{ab}{}^{cd} + 8 \d_{[a}^{[c} \frak{f}_{b]}{}^{d]} + \psi_{[a}{}^\a_I \phi_{b]}{}^{\b I} (\g_f)_{\a\b} \eps^{fcd} \ , \label{LCConst}
\ee
where $\cR_{ab}{}^{cd}$ is the Lorentz curvature
constructed solely from the spin connection,\footnote{We caution
the reader that $\cR_{ab}{}^{cd}$ does not possess the usual symmetries
of the Riemann tensor; in particular, its corresponding Ricci tensor
$\cR_{ac} = \eta^{bd} \cR_{abcd}$ is not symmetric in general.}
\be \label{eq:cRM}
\cR_{ab}{}^{cd} = 2 e_a{}^m e_b{}^n \partial_{[m} \omega_{n]}{}^{ab} - 2 \omega_{[a}{}^{cf} \omega_{b]}{}_f{}^d \ .
\ee

Contracting $b$ with $d$ in eq. \eqref{LCConst} gives
\be 0 = \cR_{ab} + 2 \frak{f}_{ab} + 2 \eta_{ab} \frak{f}^c{}_c - \hf \eps_b{}^{cd} \psi_c{}^\a_I \phi_d{}^{\b I} (\g_a)_{\a\b} + \hf \eta_{ab} \eps^{cde} \psi_c{}^\a_I \phi_d{}^{\b I} (\g_e)_{\a\b} \ ,
\ee
which may be solved for the special conformal connection
\be \frak{f}_{ab} = - \hf \cR_{ab} + \frac{1}{8} \eta_{ab} \cR + \frac{1}{4} \eps_{b}{}^{cd} (\g_a)_{\a\b} \psi_c{}^\a_I \phi_d{}^{\b I} 
- \frac{1}{8} \eta_{ab} \eps^{def} \psi_d{}^\a_I \phi_e{}^{\b I} (\g_f)_{\a\b} \ . \label{Fcomp}
\ee
As a consistency check, we note that eq. \eqref{confPhi}
and the explicit definition of the spin connection implies that
\be \frak{f}_{[ab]} = - \hf e_a{}^m e_b{}^n \partial_{[m} b_{n]} - \frac{1}{4} \psi_{[a}{}^\a_I \phi_{b]}{}_\a^I \ ,
\ee
which agrees with eq. \eqref{DilEqfF}.

The results derived so far are all we need to construct the component actions.
The analysis of the $\rm SO(\cN)$, $S$-supersymmetry and special conformal
curvatures remains. This will give explicit expressions for the composite
component fields of the super Cotton tensor but yield no new results, so we
confine that discussion to Appendix \ref{compAnalysis2}.

%%%%%%%%%%%%%%%%%%%%%%%%%%%%%%%%%%%%%%%%%%%%%%%%%%%%%%
%%%%%%%%%%%%%%%%%%%%%%%%%%%%%%%%%%%%%%%%%%%%%%%%%%%%%%

\section{Off-shell component actions} \label{actions}

In section \ref{CSCI} we constructed the appropriate closed forms
which describe the off-shell conformal supergravity actions for 
$\cN < 6$. All that remains is to make use of the component
results of the previous section to explicitly construct these actions.
This is our goal in this section.

%%%%%%%%%%%%%%%%%%%%%%%%%%%%%%%%%%%%%%%%%%%%%%%%%%%%%%

\subsection{The Chern-Simons contribution}

We first write down the Chern-Simons contribution to the action,
which has a universal form for all values of $\cN$, aside from
obvious truncations at $\cN=1$ and $\cN=2$.
It helps at this point to recall that $b_m$ is the only fundamental
field in the Chern-Simons action that transforms under $K$. As the action is
$K$-invariant up to a total derivative, it follows that the dependence
on $b_m$ must drop out \cite{RvanN86}. Equivalently, we can simply
adopt the $K$-gauge $b_m=0$. Using the action \eqref{ectoS} and the
Chern-Simons form \eqref{CSFORM}, we find the Chern-Simons contribution to be
\begin{align}
S_{\rm CS}
=& \ \frac{1}{4} \int \rd^3 x \,e \,\eps^{abc} \Big( \omega_{a}{}^{fg} \cR_{bc}{}_{fg} - \frac{2}{3} \omega_{af}{}^g \omega_{bg}{}^h \omega_{ch}{}^f \non\\
&+ 4 \ri \phi_b{}^{\a I} \phi_c{}^\b_I (\g_a)_{\a\b} - \hf {\cR}_{ab}{}^{IJ} V_c{}_{IJ} - \frac{4}{3} V_a{}^{IJ} V_b{}_I{}^K V_c{}_{KJ} \Big) \ ,
\end{align}
where $\cR_{ab}{}^{cd}$ and $\cR_{ab}{}^{IJ}$ are defined respectively in eqs. 
\eqref{eq:cRM} and \eqref{eq:cRN}.
Using equation \eqref{confPhi} for the explicit form of the $S$-supersymmetry
connection, the Chern-Simons action becomes
\begin{align}\label{eq:Scs}
S_{\rm CS} =& \ \frac{1}{4} \int \rd^3 x \,e \,\eps^{abc} \,\Big( \omega_{a}{}^{fg} \cR_{bc}{}_{fg}  - \frac{2}{3} \omega_{af}{}^g \omega_{bg}{}^h \omega_{ch}{}^f 
-  \frac{\ri}{2} \Psi_{bc}{}^\a_I (\g_d)_\a{}^\b (\g_a)_\b{}^\g \eps^{def} \Psi_{ef}{}^I_\g  \non\\
&- 2 {\cR}_{ab}{}^{IJ} V_c{}_{IJ} - \frac{4}{3} V_a{}^{IJ} V_b{}_I{}^K V_c{}_{KJ} \Big) \ .
\end{align}
This coincides with the actions studied in \cite{LR89, NG}.

%%%%%%%%%%%%%%%%%%%%%%%%%%%%%%%%%%%%%%%%%%%%%%%%%%%%%%

\subsection{The full conformal supergravity action}

For $\cN \leq 2$, the Chern-Simons action $S_{\rm {CS}}$ is the full
off-shell action for conformal supergravity \cite{vanN85, RvanN86}.
Higher values of $\cN$ require the introduction of an auxiliary field
sector, which, for the values $3 \leq \cN \leq 5$, can be described
using a curvature induced three form. Each case for $\cN<6$ is summarized below.

%%%%%%%%%%%%%%%%%%%%%%%%%%%%%%%%%%%%%%%%%%%%%%%%%%%%%%

\subsubsection{The $\cN = 1$ case}

The $\cN=1$ conformal supergravity action can be read off from eq. 
\eqref{eq:Scs} by turning off the $\rm SO(\cN)$ contribution,\footnote{The $\cN = 0$ case
can further be read off by eliminating the gravitinos.}
\begin{align}
S =& \ \frac{1}{4} \int \rd^3 x \,e \,\eps^{abc} \Big(
	\omega_{a}{}^{fg} \cR_{bc}{}_{fg}
	- \frac{2}{3} \omega_{af}{}^g \omega_{bg}{}^h \omega_{ch}{}^f
% 	\non\\&
	-  \frac{\ri}{2} \Psi_{bc}{}^\a_I (\g_d)_\a{}^\b (\g_a)_\b{}^\g \eps^{def} \Psi_{ef}{}^I_\g \Big) \ ,
\end{align}
which agrees with the action given in \cite{vanN85}.

%%%%%%%%%%%%%%%%%%%%%%%%%%%%%%%%%%%%%%%%%%%%%%%%%%%%%%

\subsubsection{The $\cN = 2$ case}

The $\cN=2$ action is similarly a direct consequence of eq. 
\eqref{eq:Scs}, keeping in mind that the group $\rm SO(2)$ is abelian:
\begin{align}
S =& \ \frac{1}{4} \int \rd^3 x \,e \,\eps^{abc} \Big( \omega_{a}{}^{fg} \cR_{bc}{}_{fg} - \frac{2}{3} \omega_{af}{}^g \omega_{bg}{}^h \omega_{ch}{}^f \non\\
&-  \frac{\ri}{2} \Psi_{bc}{}^\a_I (\g_d)_\a{}^\b (\g_a)_\b{}^\g \eps^{def} \Psi_{ef}{}^I_\g - 2 {\cR}_{ab}{}^{IJ} V_c{}_{IJ} \Big) \ .
\end{align}
This coincides with the action constructed in \cite{RvanN86}.

%%%%%%%%%%%%%%%%%%%%%%%%%%%%%%%%%%%%%%%%%%%%%%%%%%%%%%

\subsubsection{The $\cN = 3$ case}

This is the first case where auxiliary fields occur. Here there is
a single auxiliary field, the spinor $w_\alpha$ defined in \eqref{eq:CompN=3}.
Its contribution to the action can be found by taking the appropriate projection of
the curvature induced form ${\bm\S}_R$. Using the formula
\begin{align}\label{eq:SigmaProjection}
\frac{1}{3 !} \eps^{mnp} {{\bm \S}}_{mnp}| &=
	\frac{1}{3 !} \eps^{mnp} E_p{}^C E_n{}^B E_m{}^A {\bm \S}_{ABC}| \non \\
	&= \frac{1}{3 !} \eps^{abc} \big( {{\bm \S}}_{abc}| + \frac{3}{2} \psi_a{}^\a_I {{\bm \S}}_\a^I{}_{bc}| + \frac{3}{4} \psi_b{}^\b_J \psi_a{}^\a_I {{\bm \S}}_\a^I{}_\b^J{}_{c}|
	\non\\&\quad
	+ \frac{1}{8} \psi_c{}^\g_K \psi_b{}^\b_J \psi_a{}^\a_I {{\bm \S}}_\a^I{}_\b^J{}_\g^K| \big)
\end{align}
for the component projection of a three-form along with the explicit
expressions \eqref{sigmaN=3} for the components of $\bm\S_{ABC}$, we find
\begin{align}
\frac{1}{3 !} \eps^{mnp} {{\bm \S}}_{mnp}| &= 8 \ri \,W^\a| W_\a| = 8 \ri \,w^\a w_\a \ .
\end{align}
Combining this result with the Chern-Simons contribution gives the full action
\begin{align}
S =& \  \frac{1}{4} \int \rd^3 x \,e \,\Big\{
	\eps^{abc} \big( \omega_{a}{}^{fg} \cR_{bc}{}_{fg} - \frac{2}{3} \omega_{af}{}^g \omega_{bg}{}^h \omega_{ch}{}^f
		-  \frac{\ri}{2} \Psi_{bc}{}^\a_I (\g_d)_\a{}^\b (\g_a)_\b{}^\g \eps^{def} \Psi_{ef}{}^I_\g 
	\non\\ &
	- 2 {\cR}_{ab}{}^{IJ} V_c{}_{IJ} - \frac{4}{3} V_a{}^{IJ} V_b{}_I{}^K V_c{}_{KJ} \big)
	- 32 \ri \,w^\a w_\a \Big\} \ .
\end{align}

%%%%%%%%%%%%%%%%%%%%%%%%%%%%%%%%%%%%%%%%%%%%%%%%%%%%%%

\subsubsection{The $\cN = 4$ case}

For $\cN=4$, the auxiliary fields content expands to a spinor $\rm SO(4)$-vector $w_\alpha^I$
and two real scalars $w$ and $y$. Again using the expression
\eqref{eq:SigmaProjection}, now with the components \eqref{sigmaN=4} of $\bm\S_{ABC}$,
one finds
\begin{align}
\frac{1}{3 !} \eps^{mnp} {{\bm \S}}_{mnp}|
	&= - 2 \ri (\nabla^\a_I W)|(\nabla_\a^I W)| - \frac{\ri}{2}W|(\nabla^\a_I \nabla_\a^I W)|
	\non\\& \quad
	- 2 (\g^a)_{\a\b} \psi_a{}^\a_I W| \nabla^{\b I} W| + \frac{\ri}{2} \eps^{abc} (\g_a)_{\a\b} \psi_b{}^\a_I \psi_c{}^{\b I} (W|)^2 \non \\
	&= 8 \ri w^\a_I w_\a^I + 2 w y
% 	\non\\&\quad
	- 4 \ri (\g^a)_{\a\b} \psi_a{}^\a_I w w^{\b I} + \frac{\ri}{2} \eps^{abc} (\g_a)_{\a\b} \psi_b{}^\a_I \psi_c{}^{\b I} w^2 \ .
\end{align}
We have relabelled the components \eqref{defCompW} into a form appropriate for $\cN=4$:
\be \label{eq:N4auxredef}
w := \frac{1}{4!} \eps_{IJKL} w^{IJKL} \ , \quad y := \frac{1}{4!} \eps_{IJKL} y^{IJKL} \ , \quad w_{\a L} := \frac{1}{3!} \eps_{IJKL} w_\a{}^{IJK} \ .
\ee
The full $\cN=4$ conformal supergravity action is
\begin{align}
S =& \  \frac{1}{4} \int \rd^3 x \,e \,\Big\{ \eps^{abc} \big( \omega_{a}{}^{fg} \cR_{bc}{}_{fg} - \frac{2}{3} \omega_{af}{}^g \omega_{bg}{}^h \omega_{ch}{}^f 
-  \frac{\ri}{2} \Psi_{bc}{}^\a_I (\g_d)_\a{}^\b (\g_a)_\b{}^\g \eps^{def} \Psi_{ef}{}^I_\g  \non\\
&- 2 {\cR}_{ab}{}^{IJ} V_c{}_{IJ} - \frac{4}{3} V_a{}^{IJ} V_b{}_I{}^K V_c{}_{KJ} \big) \non\\
&- 32 \ri w^\a_I w_\a^I - 8 w y
	- 16 \ri \,\psi_a{}^\a_I (\g^a)_{\a}{}^\b w_\b^{I} w - 2 \ri \eps^{abc} (\g_a)_{\a\b} \psi_b{}^\a_I \psi_c{}^{\b I} w^2 \Big\} \ .
\end{align}

%%%%%%%%%%%%%%%%%%%%%%%%%%%%%%%%%%%%%%%%%%%%%%%%%%%%%%

\subsubsection{The $\cN = 5$ case}

For our final case, we need the expressions in \eqref{sigmaN=5}, which yield
\begin{align}
\frac{1}{3 !} \eps^{mnp} {{\bm \S}}_{mnp}|
	&= \frac{2 \ri}{25}(\de^{\r P}W_P)|(\de_\r^QW_Q)| - \ri (\de_{[P}^\r W_{Q]})|(\de_{\r}^{[P} W^{Q]})| \non\\
	&\quad- \frac{2\ri }{3}(\de^\r_{ P}\de_{\r}^{P}W^{S})|W_S|
	\non\\&\quad
	+ 2 (\g^a)_\a{}^\b \psi_a{}^\a_I \big(\nabla_\b^{[I} W^{J]}| W_J| - \frac{1}{5} (\nabla_\b^J W_J|) W^I| \big)
	\non\\&\quad
	+ \frac{\ri}{2} \eps^{abc} (\g_a)_{\a\b} \psi_b{}^\a_I \psi_c{}^\b_J (\d^{IJ} W^K| W_K| - 2 W^I| W^J|) \non\\
	&= - 2 \ri X^\a X_\a + 4 \ri w^{\a IJ} w_{\a IJ}
	+ 2 w^I y_I
% 	\non\\&\quad
	+ 2 (\g^a)_\a{}^\b \psi_a{}^\a_I \big(2 \ri w_\b{}^{IJ} w_J + \ri X_\b w^I \big)
	\non\\&\quad
	+ \frac{\ri}{2} \eps^{abc} (\g_a)_{\a\b} \psi_b{}^\a_I \psi_c{}^\b_J (\d^{IJ} w^K w_K - 2 w^I w^J)~.
\end{align}
For $\cN=5$ we have defined our auxiliary fields as
\begin{align}
w_I &:= \frac{1}{4!} \eps_{IJKLP} w^{JKLP} = W_I | \ , \qquad
	y_I:= \frac{1}{4!} \eps_{IJKLP} y^{JKLP} = - \frac{\ri}{3} \nabla^\g_P \nabla_\g^P W_I | \non\\
w_\a{}^{IJ} &:= \frac{1}{3!} \eps^{IJKLP} w_{\a KLP} = - \frac{\ri}{2} \nabla_\a^{[I} W^{J]}| \ ,
	\non\\
X_\a &:= \frac{1}{5!} \eps_{IJKLP} X_\a{}^{IJKLP} = \frac{\ri}{5} \nabla_\a^I W_I | \ .
\end{align}
They consist of two real bosonic $\rm SO(5)$ vectors, $w^I$ and $y^I$, as well as two spinors,
an $\rm SO(5)$ singlet, $X_\alpha$, and an antisymmetric $\rm SO(5)$ tensor, $w_\a{}^{IJ}$.

The off-shell $\cN=5$ action is
\begin{align}
S =& \  \frac{1}{4} \int \rd^3 x \,e \,\Big\{ \eps^{abc} \big( \omega_{a}{}^{fg} \cR_{bc}{}_{fg} - \frac{2}{3} \omega_{af}{}^g \omega_{bg}{}^h \omega_{ch}{}^f 
-  \frac{\ri}{2} \Psi_{bc}{}^\a_I (\g_d)_\a{}^\b (\g_a)_\b{}^\g \eps^{def} \Psi_{ef}{}^I_\g 
	\non\\&
	- 2 {\cR}_{ab}{}^{IJ} V_c{}_{IJ} - \frac{4}{3} V_a{}^{IJ} V_b{}_I{}^K V_c{}_{KJ} \big)
	\non\\&
	+ 8 \ri X^\a X_\a - 16 \ri w^{\a IJ} w_{\a IJ} - 8 w^I y_I
% 	\non\\ &
	- 8 \ri \, \psi_a{}^\a_I (\g^a)_\a{}^\b \big(2 w_\b{}^{IJ} w_J + X_\b w^I \big) \non\\
& - 2 \ri \eps^{abc} (\g_a)_{\a\b} \psi_b{}^\a_I \psi_c{}^\b_J (\d^{IJ} w^K w_K - 2 w^I w^J) \Big\} \ .
\end{align}

Our choice of normalization for the auxiliary fields allows a simple truncation
to lower values of $\cN$. Beginning with $\cN=5$, one truncates the
auxiliary fields to $\cN=4$ by taking
\begin{align}\label{eq:5to4Trunc}
\left.\begin{gathered}
w_I \longrightarrow 0~, \qquad 
w_\alpha{}^{IJ} \longrightarrow 0~, \qquad
X_\alpha \longrightarrow 0~, \qquad
y_I \longrightarrow 0~, \\
w_5 \longrightarrow w~, \qquad 
w_\alpha{}^{I5} \longrightarrow w_\alpha{}^{I} ~, \qquad
y_5 \longrightarrow y~
\end{gathered} \quad \right\} \quad I,J=1,2,3,4~.
\end{align}
One can check using the transformation rules in Appendix \ref{SUSY}
that this truncation is consistent with the
$\cN=4$ supersymmetry and $S$-supersymmetry.
Similarly, one can truncate the $\cN=4$ action to $\cN=3$ by taking
\begin{align}\label{eq:4to3Trunc}
w \longrightarrow 0~, \qquad 
w_\alpha{}^{I} \longrightarrow 0~, \qquad
w_\alpha{}^4 \longrightarrow w_\alpha ~, \qquad
y \longrightarrow 0~, \qquad
I=1,2,3~.
\end{align}
The truncation procedure for the gauge fields is obvious.

The off-shell actions for $\cN=3$, $\cN=4$ and $\cN=5$ are new actions and are the main
results of this paper.

%%%%%%%%%%%%%%%%%%%%%%%%%%%%%%%%%%%%%%%%%%%%%%%%%%%%%%
%%%%%%%%%%%%%%%%%%%%%%%%%%%%%%%%%%%%%%%%%%%%%%%%%%%%%%

\section{Conclusion} \label{conclusion}

In this paper we constructed the off-shell actions for all three-dimensional 
conformal supergravity theories with $\cN<6$, both in superspace and in terms of the component
fields. In the simplest cases $\cN=1$ and $\cN=2$, our component actions coincide
with those derived  in \cite{vanN85} and \cite{RvanN86}, respectively, using the superconformal 
tensor calculus.\footnote{The action for $\cN=1$ conformal supergravity \cite{vanN85}
is a natural reformulation of the $\cN=1$ supersymmetric Lorentz Chern-Simons term 
\cite{DK,Deser}.} To the best of our knowledge, 
the off-shell actions for $\cN=3, 4, 5$ conformal supergravity theories are new results. 
In the $\cN=4$ case, only the linearized conformal supergravity action was known before
\cite{Bergshoeff:2010ui}. 

Our analysis was based on the use of conformal superspace \cite{BKNT-M1},  which is a new 
formulation for the $\cN$-extended conformal supergravity. As compared with the 
conventional formulation developed earlier \cite{HIPT,KLT-M11}, 
conformal superspace is much more efficient as far as the conformal supergravity actions 
are concerned. To appreciate the power of the approach of \cite{BKNT-M1}, 
it suffices to compare the 
off-shell constructions of the $\cN=1$ conformal supergravity action given 
in this paper using conformal superspace 
and in \cite{KT-M12}  using the conventional formalism. 

Although we successfully constructed the conformal supergravity actions
for all $\cN<6$, overcoming the $\cN = 6$ barrier still remains a very interesting problem. 
We remind the reader that at the heart of our construction are two fundamental ingredients:
(i) the Chern-Simons three-form ${\bm \S}_{\rm CS}$, constructed in subsection 4.1,  
which is well defined for any $\cN$;
and (ii) the curvature induced form ${\bm \S}_{ R}$ that we constructed for $\cN<6$ in subsection 4.2.
The closed three-form ${\bm{\frak{J}}}={\bm \S}_{\rm CS}-{\bm \S}_{ R}$ was then used to build 
the conformal supergravity actions. 
A natural question is why the general ansatz \eqref{ansatz} did not work for $\cN\geq 6$.
One possibility is that for $\cN\geq 6$ it is necessary to impose an extra constraint on the super Cotton tensor
setting eq. \eqref{4.28} to zero.
Another possibility is that for $\cN\geq 6$ the ansatz \eqref{ansatz} has to be extended.
Both possibilities are ultimately related to the existence of the sequence of fields 
$X_{\a_1\cdots\a_n}{}^{I_1\cdots I_{n+4}} $ in the Weyl multiplet, which 
appear on the left hand side of Figure 1. It can be proved that all these fields with $n>1$
satisfy, in the linearized approximation,
the conservation equations 
$(\g^a)^{\b\g}\pa_aX_{\b\g\a_3\cdots\a_n}{}^{I_1\cdots I_{n+4}}=0$.\footnote{To the best of 
our knowledge a full non-linear extension of this equation has never appeared in the literature for general $\cN$.}
This indicates that the superfields $X_{\a_1\cdots\a_n}{}^{I_1\cdots I_{n+4}}$ with $n>1$ are composite
field strengths of hidden (super)symmetries (see e.\,g.~\cite{Bergshoeff:2010ui}).
It then seems clear that a possible way of addressing the $\cN\geq 6$ case is by either
finding a consistent way to truncate part of the Weyl multiplet fields, or by adding the potentials 
of the field strengths $X_{\a_1\cdots\a_n}{}^{I_1\cdots I_{n+4}}$, $n>1$, in a properly extended 
ansatz for ${\bm \S}_{ R}$.
We hope  to address these issues in the future.

The results of this paper allow one to construct off-shell actions for $\cN\leq 4$ topologically 
massive supergravity. In general, such an action is given as a sum of three terms: 
(i) the Poincar\'e supergravity action; (ii) the locally supersymmetric cosmological term; 
and (iii) the conformal supergravity action.  
The off-shell action for $\cN=1$ topologically massive supergravity was first given in 
\cite{RvanN86}, with the building blocks (i) and (ii) taken from \cite{GGRS}.
Upon elimination of the auxiliary scalar, 
this action reduces to that originally given in \cite{DK,Deser}.
In the case $\cN=2$, the superspace building blocks (i) and (ii) are given in \cite{KLT-M11,KT-M11}.
The interesting feature of this case is that there exist several off-shell versions for $\cN=2$ 
Poincar\'e supergravity, which lead to different topologically 
massive supergravity theories. These theories will be studied in \cite{KLRST-M}.
In the cases $\cN=3$ and $\cN=4$,  the superspace building blocks (i) and (ii) are given 
in \cite{KLT-M11}. 

Since the off-shell action for $\cN=5$ conformal supergravity has been given in this paper,
an interesting open problem is to develop an off-shell formulation for $\cN=5$ Poincar\'e supergravity. 
In the rigid supersymmetric case, Zupnik has derived, 
building on the earlier work by Howe and Leeming \cite{HL},
 harmonic-superspace 
formulations for the $\cN=5$ vector multiplet and corresponding Chern-Simons actions
\cite{Z2007,Z2008}. However, to the best of our knowledge, 
no off-shell results are yet available for $\cN=5$ Poincar\'e (or anti-de Sitter) supergravity. 

To construct the off-shell conformal supergravity actions, we made use of the 
%ectoplasm formalism
superform approach
 for the construction of supersymmetric invariants\footnote{One of the first applications of the formalism presented in  \cite{Castellani}
was given by Bandos, Sorokin and 
Volkov \cite{BSV}.}
 \cite{Castellani,Hasler,Ectoplasm,GGKS} in the presence of a Chern-Simons term. 
This is  an example of a known construction where an invariant derived from a closed super
$d$-form can be generated from a closed, gauge-invariant super $(d+1)$-form
provided that the latter is Weil trivial, i.e. exact in invariant cohomology
(a concept introduced by Bonora, Pasti and Tonin \cite{BPT} in the
context of anomalies in supersymmetric theories). Examples of this include
Green-Schwarz actions for various branes \cite{HRS}, as well as some
higher-order invariants in other supersymmetric theories which were studied, e.g.,
 in \cite{BHLSW,BHS13}.
\\

%%%%%%%%%%%%%%%%%%%%%%%%%%%%%%%%%%%%%%%%%%%%%%%%%%%%%%
%%%%%%%%%%%%%%%%%%%%%%%%%%%%%%%%%%%%%%%%%%%%%%%%%%%%%%

\noindent
{\bf Acknowledgements:}\\
SMK thanks Dima Sorokin for pointing out the important work \cite{Castellani}.
We are grateful to the referee of this paper for useful suggestions and for bringing 
the references \cite{BPT,HRS} to our attention. 
The work of DB was supported by
ERC Advanced Grant No. 246974, ``{\it Supersymmetry: a window to
non-perturbative physics}.''
The work of SMK  and JN was supported in part by the Australian Research Council,
project No. DP1096372.  
The work of GT-M and JN was supported by the Australian Research Council's Discovery Early Career 
Award (DECRA), project No. DE120101498.

%%%%%%%%%%%%%%%%%%%%%%%%%%%%%%%%%%%%%%%%%%%%%%%%%%%%%%
%%%%%%%%%%%%%%%%%%%%%%%%%%%%%%%%%%%%%%%%%%%%%%%%%%%%%%

\appendix

\section{Note on gauge invariant forms} \label{InvForms}

In this appendix we collect some useful notes for checking the gauge invariance of superforms.

First let $\S$ be some gauge invariant $p$-form (under $\cH$)
\be \d_\cH \S = 0 \ , \quad \S = \frac{1}{p !} E^{A_p} \wedge \cdots \wedge E^{A_1} \S_{A_1 \dots A_p} \ .
\ee
We recall that the gauge transformation of the vielbein (under $\cH$) is
\be \d_\cH E^A = E^B \L^{\underline{c}} f_{\underline{c} B}{}^A \ .
\ee
Then since $\S$ is invariant we require
\be \d_\cH \S_{A_1 \cdots A_p} = - p \L^{\underline{a}} f_{\underline{a} [A_1}{}^D \S_{|D|A_2 \cdots A_p\}} \ ,
\ee
or, equivalently,
\be X_{\underline{a}} \S_{A_1 \dots A_p} = - f_{\underline{a} [A_1}{}^D \S_{|D|A_2\cdots A_p \}} \ , \quad \d_\cH \S_{A_1 \cdots A_p} = \L^{\underline{a}} X_{\underline{a}} \S_{A_1 \cdots A_p} \ .
\ee
From here it is easy to see that under Lorentz and $\rm SO(\cN)$ transformations we require $\S_{A_1 \cdots A_p}$ to transform as a tensor. 
Furthermore, using the superconformal algebra we find that the dimension of each of the component fields is given by
\be \mathbb D \S_{A_1 \cdots A_p} = (\D_{A_1} + \cdots + \D_{A_p}) \S_{A_1 \cdots A_p} \ ,
\ee
where $\D_A$ is the dimension of $P_A$,
\be [\mathbb D, P_A] = \D_A P_A \ .
\ee

The remaining gauge transformations are the special conformal transformations. Now since the vielbein does not transform under the special 
conformal boosts we actually find that $\S_{A_1 \cdots A_p}$ is annihilated by $K_a$,
\be K_a \S_{A_1 \dots A_p} = - f_{K_a [A_1}{}^D \S_{|D|A_2 \cdots A_p \}} = 0 \ .
\ee

The $S$-supersymmetry transformation of $\S_{A_1 \cdots A_p}$ is given by
\be S_\a^I \S_{A_1 \cdots A_p} = - p f_{S_\a^I [A_1}{}^D \S_{|D|A_2 \cdots A_p\}} \ .
\ee
Its consequences are less trivial. Using the superconformal algebra we find
\be S_\a^I \S_{A_1 \cdots A_p} = \ri p (\g_a)_\a{}^\b \S_\b^I{}_{[A_2\cdots A_p} \d_{A_1\}}^a \ , \label{1rec}
\ee
which leads to the relation
\be S_\b^J \S_{a_1 \cdots a_{n}}{}_{\a_1}^{I_1}{\cdots}{}_{\a_{p - n}}^{I_{p-n}} = 
\ri n (\g_{[a_{1}})_\b{}^\g \S_\g^J{}_{a_2 \cdots a_{n}]}{}_{\a_1}^{I_1}{\cdots}{}_{\a_{p - n}}^{I_{p-n}}{}  \ . \label{recurs}
\ee
Equation \eqref{1rec} automatically implies that the lowest non-zero component of $\S$ is primary.

%%%%%%%%%%%%%%%%%%%%%%%%%%%%%%%%%%%%%%%%%%%%%%%%%%%%%%
%%%%%%%%%%%%%%%%%%%%%%%%%%%%%%%%%%%%%%%%%%%%%%%%%%%%%%

\section{Matrix realization of the $\cN$-extended superconformal algebra} \label{matrixReal}

In this paper we made use of the adjoint representation of the superconformal algebra \eqref{SCA}. 
In some situations, however, 
it is more advantageous  to use instead 
its fundamental representation. 
The important advantage of the latter is that the Cartan-Killing metric, in general, can only be 
computed in such a representation, since the index of the adjoint representation may vanish for certain simple supergroups \cite{Kac,DeWitt}.\footnote{This is precisely what 
we observed in subsection \ref{subsec4.1} for  the $\cN = 6$ case.} 
In this appendix, we give the matrix realization of the superconformal group that was described, e.g.,  in \cite{KPT-MvU}
and use it to compute the Cartan-Killing metric.

In order to 
describe the fundamental representation of 
the $\cN$-extended superconformal algebra\footnote{Ref. \cite{KPT-MvU} followed 
slightly different conventions by denoting the superalgebra by ${\mathfrak{osp}}(\cN|2, {\mathbb R})$.}  
$\mathfrak{osp}(\cN|4, {\mathbb R})$
we introduce the symplectic supermetric
\bea
{\mathbb J} =  \left(
\begin{array}{cc}
J ~&~ 0 \\
0 ~& ~{\rm i} \,{\mathbbm 1}_\cN 
\end{array} \right) \ , \label{supermetric}
\eea
where $J$ is the skew-symmetric supermatrix 
\be J = \big(J^{\hat \a \hat \b} \big) 
=\left(
\begin{array}{cc}
0  & {\mathbbm 1}_2 \\
 -{\mathbbm 1}_2  &    0
\end{array}
\right) ~.
\ee
One may naturally associate with the symplectic supermetric $\mathbb J$ 
the quadratic form on ${\mathbb R}^{\cN |4}$
\bea
{ \S}^{\rm sT}\, {\mathbb J}\, { \S} =\z^{\rm T}  J  \z + {\rm i}\, y^{\rm T} y ~,
\label{4.2}
\eea
which is symmetric and purely imaginary.\footnote{The quadratic form \eqref{4.2} can naturally be extended to the symmetric inner product 
on ${\mathbb R}^{\cN |4}$ defined by $
\langle { \S}| { \X} \rangle_{\mathbb J}: = { \S}^{\rm sT}{\mathbb J} \, { \X}
= \langle { \X}| { \S} \rangle_{\mathbb J}$, 
with $ \S$ and $ \X$ being arbitrary odd supertwistors \cite{KPT-MvU}.} 
The superspace ${\mathbb R}^{\cN |4}$ is 
parametrized by 4 anti-commuting real variables $\z$ and 
$\cN$ commuting real variables $y$,
\bea
{ \S} = \left(
\begin{array}{c}
 \z \\
 y 
\end{array}
\right)~, \qquad { \S}^{\rm sT} =\big( \z^{\rm T}~, ~y^{\rm T}\big) = \S^{\rm T}~, 
\qquad \e (\z) =1~, \quad \e(y) =0~,
\label{4.3}
\eea
where $\e (s) $ denotes the Grassmann parity of a supernumber $s$. Elements of the above form (\ref{4.3}) 
are called {\it odd real supertwistors}.

The supergroup $\rm{OSp}(\cN|4, {\mathbb R})$ is the group of linear transformations
\bea
z \to z' = g \, z~, \qquad
g=\left(
\begin{array}{c||c}
 A  & B\\
 \hline \hline
C &    D 
\end{array}
\right)
\label{3.4}
\eea
that leave the quadratic form (\ref{4.2}) invariant. The corresponding supermatrix $g$ obeys the equation
\bea
g^{\rm sT} {\mathbb J}\, g = {\mathbb J} ~, \qquad 
g^{\rm sT}=\left(
\begin{array}{c||r}
 A^{\rm T}  & -C^{\rm T}\\
\hline \hline
B^{\rm T}  &    D^{\rm T} 
\end{array}
\right) ~,
\eea
which gives us a supermatrix realization of $\rm{OSp}(\cN|4, {\mathbb R})$. 
The even matrices $A, D$ and the odd 
matrix $B$ in (\ref{3.4}) have real matrix elements, 
while the odd matrix $C$ has purely imaginary matrix elements. We call the
supermatrices $g$ of this type {\it real}.

The superconformal algebra $\mathfrak{osp}(\cN|4, {\mathbb R})$ consists of real supermatrices 
obeying the master equation
\bea
{ \O}^{\rm sT} {\mathbb J} + {\mathbb J} \,{ \O}=0~. \label{msla}
\eea
The general solution of eq. \eqref{msla} is\footnote{The normalization of the parameters have been chosen such 
that $\l_\a{}^\b \rightarrow \hf \l_\a{}^\b$, $b_a \rightarrow - b_a$ and $\eta^\a_I \rightarrow \ri \eta^\a_I$ relative to \cite{KPT-MvU}.}
\bea
{ \O} 
&=& \left(
\begin{array}{c | c ||c}
  \hf \l -\hf f {\mathbbm 1}_2  ~& ~ - \check{b} &~\ri \sqrt{2} \eta^{\rm T}   \\
\hline 
-\hat{a}  ~&   -  \hf \l^{\rm T} +\hf f {\mathbbm 1}_2~&~ -\sqrt{2}\e^{\rm T}
\\
\hline
\hline
{\rm i}\sqrt{2}\,\e ~& ~ - \sqrt{2}\, \eta &~r
\end{array}
\right) \non \\
&\equiv& \left(
\begin{array}{c | c ||c}
  \hf \l_\a{}^\b -\hf f \d_\a{}^\b  ~& ~ - b_{\a \b} &~\ri \sqrt{2} \eta_{\a J}   \\
\hline 
-a^{\a \b}  ~&   - \hf \l^\a{}_\b +\hf f \d^\a{}_\b~&~ -\sqrt{2}\e^\a{}_J
\\
\hline
\hline
{\rm i}\sqrt{2}\,{\e_I{}^\b} ~& ~ - \sqrt{2}\, \eta_{I\b}&~r_{IJ}
\end{array}
\right) ~, 
\label{SP-g} \\
&&  \l_\a{}^\a =0~, \qquad {a}^{\a\b} = {a}^{\b \a} ~, \qquad {b}_{\a \b} = {b}_{\b \a}~, \qquad
r_{IJ}=-r_{JI}~,
\non
\eea
with  $I,J =1,\dots, \cN$.
Here the bosonic  parameters $\l_\a{}^\b$, $f$, $a_{\a\b}$, $b^{\a\b}$ and $r_{IJ}$ are real, while the fermonic parameters 
$\e^\a{}_I\equiv \e_I{}^\a $ and $\eta_{\a I} \equiv \eta_{I\a}$ are real and imaginary respectively.

The Cartan-Killing metric on $\mathfrak{osp}(\cN|4, {\mathbb R})$ is defined by
\bea 
{\bm\G}( \O, \hat{ \O}) := - {\rm sTr}( \O \hat{ \O}) 
= - {\hat{{ \O}}}{}^{\tilde{b}} { \O}^{\tilde{a}} 
\G_{\tilde{a}\tilde{b}} \ , \quad { \O}, \hat{ \O}  \in 
\mathfrak{osp}(\cN|4, {\mathbb R}) \ ,
\eea
where $ \O =  \O^{\tilde{a}} X_{\tilde{a}}$ 
in some basis.\footnote{In such a basis, the components of the Cartan-Killing metric are 
$\G_{\tilde{a} \tilde{b}} = {\rm sTr}(X_{\tilde{a}} X_{\tilde{b}})$.} 
We may then choose a basis to correspond to the parametrization \eqref{SP-g},
\begin{align}  \O = -\hf a^{\a\b} P_{\a\b} + \hf \l^{\a\b} M_{\a\b} + \hf r^{IJ} N_{IJ} + f \mathbb D - \hf b^{\a\b} K_{\a\b} + \eps^\a_I Q_\a^I + \eta^\a_J S_\a^J \ ,
\end{align}
where the generators ($P$, $M$, $N$, $K$ and $S$) are to be understood as matrices. Using the matrix realization \eqref{SP-g}, we can 
explicitly compute the Cartan-Killing metric 
\be 
\bm \G( \O, \hat{ \O}) = - \hf \l_{ab} \hat{\l}{}^{ab} 
+ f \hat{f} - 2 b_a \hat{a}{}^a - 2 a^a \hat{b}_a - 4 \eta_{\a J} \hat{\eps}{}_J{}^\a 
+ 4 \eps^\a{}_J \hat{\eta}_{J \a} - r_{IJ} \hat{r}{}^{IJ} \ , \label{killInv}
\ee
where we have introduced, in line with our usual conventions, parameters with vector indices, {\it e.g.}
\be \l_{ab} = \hf \eps_{abc} (\g^c)^{\a\b} \l_{\a\b} \ , \quad b_a = -\hf (\g_a)^{\a\b} b_{\a\b} \ .
\ee
The components of the Cartan-Killing metric are then simply read off of eq. \eqref{killInv}. 
One can see that the Killing components are proportional to the ones computed using the adjoint representation for the non-vanishing cases ({\it i.e.} $\cN \neq 6$), eq. \eqref{KillingComps}.

%%%%%%%%%%%%%%%%%%%%%%%%%%%%%%%%%%%%%%%%%%%%%%%%%%%%%%
%%%%%%%%%%%%%%%%%%%%%%%%%%%%%%%%%%%%%%%%%%%%%%%%%%%%%%

\section{Component analysis of composite curvatures} \label{compAnalysis2}

In this appendix we complete the component analysis of the curvatures associated with the $\rm SO(\cN)$, $S$-supersymmetry and 
special conformal generators.

\subsection{$\rm SO(\cN)$ curvature}

The $\rm SO(\cN)$ curvature is given by eq. \eqref{NCurv}.
Taking its restriction to the bosonic manifold $\cM^3$ leads to the equation
\be
\RN_{mn}{}^{IJ}| = {\cR}_{mn}{}^{IJ} + 2 \psi_{[m}{}^{\a [I} \phi_{n]}{}_\a^{J]} \ ,
\ee
where for convenience we have introduced the non-conformal $\rm SO(\cN)$ curvature,
\be \label{eq:cRN}
{\cR}_{mn}{}^{IJ} := 2 \partial_{[m} V_{n]}{}^{IJ} - 2 V_{[m}{}{}^{IK} V_{n] K}{}^J \ .
\ee
The covariantized form of the curvature, which is given by $\RN_{ab}{}^{IJ}|$,
may be constructed by taking the lowest component of
\begin{align}
\RN_{mn}{}^{IJ} = E_m{}^A E_n{}^B \RN_{AB}{}^{IJ} (-1)^{\eps_A \eps_B} \ .
\end{align}
and solving for $\RN_{ab}{}^{IJ}|$. We will need some intermediate results.
For general $\cN$, the superspace components of the $\rm SO(\cN)$ curvature are
\bsubeq
\begin{align}
\RN_\a^I{}_\b^J{}^{KL} &= - 2 \ri \eps_{\a\b} W^{IJKL} \ , \\
\RN_a{}_\b^J{}^{KL} &= - \frac{1}{\cN - 3} (\g_a)_{\b\g} \nabla^\g_I W^{JIKL} \ , \\
\RN_{ab}{}^{KL} &= - \frac{\ri}{2 (\cN - 2) (\cN - 3)} \eps_{abc} (\g^c)_{\a\b} \nabla^\a_I \nabla^\b_J W^{IJKL} \ ,
\end{align}
\esubeq
with lowest components respectively given by
\begin{subequations}
\begin{align}
\RN_\a^I{}_\b^J{}^{KL}| &= - 2 \ri \eps_{\a\b} w^{IJKL} \ , \qquad
\RN_a{}_\b^J{}^{KL}| = -2 \ri (\gamma_a)_{\b\g} w^{\g JKL} \ , \\
\RN_{ab}{}^{KL}| &= -\eps_{abc} (\g^c)_{\a\b} w^{\a\b KL} \ . \label{eq:superNCurv}
\end{align}
\end{subequations}
This leads to
\begin{align} e_a{}^m e_b{}^n \RN_{mn}{}^{KL}| =
	- \eps_{abc} (\g^c)_{\a\b} w^{\a\b KL}
	- \frac{\ri}{2} (\g_{[a})_{\a\b} \psi_{b]}{}^\a_J w^{\b JKL}
	+ \frac{\ri}{2} \psi_{[a}{}^\a_I \psi_{b]}{}_{\a J} w^{IJKL} \ ,
\end{align}
from which we can determine $w_{\a\b}{}^{IJ}$:
\begin{align} w_{\a\b}{}^{IJ} =& - \frac{1}{4} \eps^{abc} (\g_a)_{\a\b} {\cR}_{bc}{}^{IJ} - \hf \eps^{abc} (\g_a)_{\a\b} \psi_{b}{}^{\g [I} \phi_{c}{}^{J]}_\g 
+ \frac{\ri}{2} \eps^{abc} (\g_a)_{\g\d} (\g_b)_{\a\b} \psi_c{}^\g_K w^{\d}{}^{IJK} \non\\
&+ \frac{\ri}{8} \eps^{abc} (\g_a)_{\a\b} \psi_b{}^\g_K \psi_c{}_{\g L} w^{IJKL} \ . \label{WalphbetaIJ}
\end{align}
The composite field $w_{\a\b}{}^{IJ}$ is equivalent to the supercovariant
$\rm SO(\cN)$ curvature \eqref{eq:superNCurv}, which can be written
\begin{align}
\RN_{ab}{}^{IJ}\vert &= \cR_{ab}{}^{IJ} + 2 \psi_{[a}{}^{\a [I} \phi_{b]}{}_\a^{J]}
	+ 2\ri \,\psi_{[a}{}^{\beta}_K (\gamma_{b]})_\beta{}^\gamma w_{\gamma}{}^{KIJ}
	- \frac{\ri}{2} \psi_a{}^{\beta}_K \psi_{b \beta L} w^{KLIJ}~.
\end{align}

%%%%%%%%%%%%%%%%%%%%%%%%%%%%%%%%%%%%%%%%%%%%%%%%%%%%%%

\subsection{$S$-supersymmetry curvature}

The definition of $S$-supersymmetry curvature \eqref{SCurv} implies
\be \RS_{mn}{}^\a_I | = \cD_{[m} \phi_{n]}{}^\a_I + \ri \psi_{[m}{}^\b_I \frak{f}_{n]}{}^a (\g_a)_\b{}^\a \ ,
\ee
where
\be \cD_{[m} \phi_{n]}{}^\a_I = \partial_{[m} \phi_{n]}{}^\a_I + \hf \omega_{[m}{}^a \phi_{n]}{}^\b_I (\g_a)_\b{}^\a + V_{[m}{}_{JI} \phi_{n]}{}^{\a J} - \hf b_{[m} \phi_{n]}{}^\a_I \ .
\ee
The covariantized form may be constructed from
\begin{align} \RS_{mn}{}^\a_I = E_m{}^A E_n{}^B \RS_{AB}{}^\a_I (-1)^{\veps_A \veps_B} \ .
\end{align}
We require the superspace curvatures
\bsubeq
\begin{align}
\RS_\a^I{}_\b^J{}_K^\g &= - \frac{\ri}{\cN - 3} \eps_{\a\b} \nabla^\g_L W^{IJKL} \ , \\
\RS_a{}_\b^J{}^\a_I &= \frac{1}{2 (\cN - 2) (\cN - 3)} (\g_a)_{\b\g} \nabla^\g_L \nabla^\a_P W^{JILP} \ , \\
\RS_{ab}{}^\a_I &= - \frac{\ri}{4 (\cN - 1) (\cN - 2) (\cN - 3)} \eps_{abc} (\g^c)_{\b\g} \nabla^\b_J \nabla^\g_K \nabla^\a_L W^{IJKL} \ ,
\end{align}
\esubeq
with the lowest components
\bsubeq
\begin{align}
\RS_\a^I{}_\b^J{}_K^\g| &= 2 \eps_{\a\b} w^{\g IJ}{}_K \ , \qquad
\RS_a{}_\b^J{}^\a_I| = - \ri \,(\g_a)_{\b\g} w^{\g \a J}{}_I \ , \\
\RS_{ab}{}^\a_I| &= - \frac{1}{4} \eps_{abc} (\g^c)_{\b\g} w^{\b\g \a I} \ .\label{eq:superScurv}
\end{align}
\esubeq
The last of these is the supercovariant $S$-supersymmetry curvature, also known
as the supercovariant Cottino tensor, which can be written
\begin{align}
\RS_{ab}{}^\a_I|
	=& \ e_{[a}{}^m e_{b]}{}^n \cD_m \phi_n{}^\a_I + \ri \,\psi_{[a}{}^\b_I \mathfrak f_{b]}{}^c (\g_c)_\b{}^\a
	- \ri \,\psi_{[a}{}^\beta_J (\gamma_{b]})_{\beta \gamma} w^{\gamma \alpha JI}
	+ \frac{1}{2} \psi_a{}^{\beta}_J \psi_{b \beta K} w^{\alpha JKI} \non \\
	=& \ e_{[a}{}^m e_{b]}{}^n \cD_m \phi_n{}^\a_I + \ri \,\psi_{[a}{}^\b_I \mathfrak f_{b]}{}^c (\g_c)_\b{}^\a
	- \frac{\ri}{2} (\psi_{[a J} \gamma_{b]} \gamma_c)^\alpha \RN^{c JI}| \non\\
	&+ \frac{1}{2} (\psi_{a J} \psi_{b K}) w^{\alpha JKI}~.
\end{align}
By using eq. \eqref{eq:superScurv}, this can be equivalently written as
\begin{align} w_{\a\b\g}{}^I
=& - \eps^{abc} e_a{}^m e_b{}^n \cD_m \phi_n{}_{\a I} (\g_c)_{\b\g }
- \ri \,\eps^{abc} \psi_a{}^\d_I {\frak f}_b{}^d (\g_d)_{\d \a} (\g_c)_{\b\g } \non\\
&- \ri \,\eps^{abc} (\g_a)_{\d}{}^\r \psi_b{}^\d_J w_{\r\a}{}^{IJ} (\g_c)_{\b\g } - \frac{1}{2} \eps^{abc} \psi_a{}^\d_J \psi_b{}_{\d K} w_{\a}{}^{IJK} (\g_c)_{\b\g} \ . \label{alphbetgamWI}
\end{align}

Using the previously derived expressions for the composite fields (eqs. \eqref{confPhi}, \eqref{LCConst} and \eqref{WalphbetaIJ}) one can check 
that the right hand side is totally symmetric in spinor indices.
This can be attributed to the Bianchi identity
\be \nabla_{[a} \nabla_{b} \nabla_{c]} = 0 \implies \eps^{abc} (\g_{a})_{\a\b} \RS_{bc}{}^\a_I = 0 ~,
\ee
where we used the fact that the torsion and dilatation and Lorentz curvatures vanish in the $[\nabla_a , \nabla_b]$ commutator.

%%%%%%%%%%%%%%%%%%%%%%%%%%%%%%%%%%%%%%%%%%%%%%%%%%%%%%

\subsection{Special conformal curvature}

Making use of eq. \eqref{KCurv}, we find
\be
\RK_{mn}{}^a | = 2 \cD_{[m} {\frak f}_{n]}{}^a - \frac{\ri}{2} \phi_{[m}{}^\a_I \phi_{n]}{}^{\b I} (\g^a)_{\a\b} \ ,
\ee
where
\be \cD_{[m} {\frak f}_{n]}{}^a= \partial_{[m} {\frak f}_{n]}{}^a - \omega_{[m}{}^{ab} {\frak f}_{n]b} - b_{[m} {\frak f}_{n]}{}^a \ .
\ee
The covariantized form may be constructed from
\begin{align} \RK_{mn}{}^c = E_m{}^A E_n{}^B \RK_{AB}{}^c (-1)^{\veps_A \veps_B} \ ,
\end{align}
using the superspace curvature components
\bsubeq
\begin{align}
\RK_\a^I{}_\b^J{}^c &= - \frac{1}{2 (\cN - 2) (\cN - 3)} \eps_{\a\b} (\g^c)^{\g\d} \nabla_{\g K} \nabla_{\d L} W^{IJKL} \ , \\
\RK_a{}_\b^J{}^c &= \frac{\ri}{4 (\cN - 1) (\cN - 2) (\cN -3)} (\g_a)_{\b\g} (\g^c)_{\d\r} \nabla^\g_K \nabla^\d_L \nabla^\r_P W^{JKLP} \ , \\
\RK_{ab}{}^c &= - \frac{1}{8 \cN (\cN - 1) (\cN - 2) (\cN - 3)} \eps_{abd} (\g^d)_{\a\b} (\g^c)_{\g\d} \nabla^\a_I \nabla^\b_J \nabla^\g_K \nabla^\d_L W^{IJKL} \ ,
\end{align}
\esubeq
which possess the projections
\begin{subequations}
\begin{align}
\RK_\a^I{}_\b^J{}^c| &= - \ri\, \eps_{\a\b} (\g^c)^{\g\d} w_{\g\d}{}^{IJ} ~, \qquad
\RK_a{}_\b^J{}^c| = \frac{1}{4} (\g_a)_{\b\g} (\g^c)_{\d\r} w^{\g \d \r J} ~, \\
\RK_{ab}{}^c| &= \frac{1}{8} \eps_{abd} (\g^d)_{\a\b} (\g^c)_{\g\d} \,w^{\a\b\d\g} \ . \label{eq:superKCurv}
\end{align}
\end{subequations}
The last of these is the supercovariant $K$-curvature (equivalently the
supercovariant Cotton tensor) and is given by
\begin{align} w_{\a\b\g\d} =& - 2 \eps^{abc} (\g_a)_{\a\b} e_b{}^m e_c{}^n \cD_m {\frak f}_n{}^d (\g_d)_{\g\d} - \ri \eps^{abc} (\g_a)_{\a\b} \phi_{b}{}^I_{(\g} \phi_c{}_{\d ) I} \non\\
&- \frac{1}{2} \eps^{abc} (\g_a)_{\s}{}^\r (\g_b)_{\a\b} \psi_c{}^\s_J w_{\g\d\r}{}^J + \frac{\ri}{2} \eps^{abc} (\g_a)_{\a\b} \psi_b{}^\r_J \psi_c{}_{\r K} w_{\g\d}{}^{JK} \ .
\label{eq:Cotton1}
\end{align}
Using eq. \eqref{eq:superKCurv}, we find
\begin{align}
\RK_{ab}{}^c| &= 2 \,e_a{}^m e_b{}^n \cD_{[m} {\frak f}_{n]}{}^c
	+ \frac{\ri}{2} (\phi_{[a}{}^I \gamma^c \phi_{b] I})
	\non \\ & \quad
	- (\psi_{[a}^J \gamma_{b]})^\beta \RS^c{}_{\beta J}|
	+ \frac{\ri}{4} (\psi_a^J \psi_b^K) \RN^c{}_{JK}|~.
\label{eq:Cotton2}
\end{align}

Using the expressions \eqref{Fcomp} and \eqref{alphbetgamWI}
for the component fields,
one can show that the right hand side of eq. \eqref{eq:Cotton1}
is completely symmetric in its spinor indices.
This symmetry may be attributed to the Bianchi identity
\be \nabla_{[a} \nabla_b \nabla_{c]} = 0 \implies \RK_{ab}{}^b = \RK_{[abc]} = 0 \ .
\ee

%%%%%%%%%%%%%%%%%%%%%%%%%%%%%%%%%%%%%%%%%%%%%%%%%%%%%%
%%%%%%%%%%%%%%%%%%%%%%%%%%%%%%%%%%%%%%%%%%%%%%%%%%%%%%

\section{The supersymmetry transformations} \label{SUSY}

Here we present the complete supersymmetry and $S$ transformations for
the component fields of the Weyl multiplet for $\cN<6$.

We begin with the component gauge connections defined in eqs. 
\eqref{eq:GaugeFields} and \eqref{eq:compGaugeFields}.
Because a supersymmetry transformation corresponds to a
covariant diffeomorphism $\d_{\rm cgct}(\xi)$ with parameter
$\xi^A = (0 , \xi^\a_I)$, we may immediately derive,
using eq. \eqref{eq:deltaGConn}, that
\begin{subequations}
\begin{align}
\delta_Q(\xi) e_m{}^a &= \ri (\psi_m^I \g^a \xi_I) \ , \\
\delta_Q(\xi) \psi_m{}^{\alpha I} &=  2\cD_m \xi^{\alpha I} =
	2 \pa_m \xi^{\a I} + \omega_m{}^\a{}_\b \xi^{\b I}
	+ b_m \xi^{\a I}
	- 2 V_m{}^I{}_J \xi^{\a J}, \\
\delta_Q(\xi) b_m &=
	-\phi_m^J \xi_J \ , \\
\delta_Q(\xi) V_m{}^{IJ} &= 2 \phi_m{}^{[I} \xi^{J]} - \ri (\psi_m{}^{K} \xi^L) w^{IJKL}
	- 2\ri (\xi_K \gamma_m)^\a w_\a{}^{IJK} \ , \label{eq:dQphiN} \\
\delta_Q(\xi) \omega_m{}^{ab} &=
	- \veps^{abc} (\phi_m^I \gamma_c \xi_I) \ , \\
\delta_Q(\xi) \phi_m{}^{\alpha I} &=
	-2 \ri (\xi^I \gamma_b)^\a \frak{f}_m{}^b
	+ 2 (\psi_{m}^J \xi^K) w^{\a I J K}
	+ \ri (\xi^J \g_m \g_c)^\a \RN^{c J I}| \ , \\
\delta_Q(\xi) \frak{f}_m{}^a &= \frac{\ri}{2} (\psi_m^I \xi^J) \RN^{a IJ}|
	+ \xi_I \gamma_m \RS^{a}{}^I| \ .
\end{align}
\end{subequations}
In the transformation law of the gravitino, we have made convenient use of
the covariant derivative
\be
\cD_a = e_a{}^m \cD_m = e_a{}^m (\partial_m - \hf \omega_m{}^{bc} M_{bc} - \hf V_m{}^{IJ} N_{IJ} - b_m \mathbb D) \ .
\ee
The $S$-supersymmetry transformations follow from \eqref{TConnec} with $\L^{A} = (0 , \eta^\a_I)$,
\begin{subequations}
\begin{align}
\delta_S(\eta) e_m{}^a &= 0~, \\
\delta_S(\eta) \psi_m{}^{\alpha I} &= - 2\ri \,(\eta^I \gamma_m)^\a \ , \\
\delta_S(\eta) b_m &= \psi_m^I \eta_I \ , \\
\delta_S(\eta) V_m{}^{IJ} &= 2 (\psi_m^{[I} \eta^{J]}) \ , \\
\delta_S(\eta) \omega_m{}^{ab} &= -\veps^{abc} (\psi_m^I \gamma_c \eta_I) \ , \\
\delta_S(\eta) \phi_m{}^{\alpha I} &=
	2 \cD_m \eta^{\alpha I} =
	2 \pa_m \eta^{\alpha I} + \omega_m{}^{\alpha}{}_\beta \eta^{\beta I}
	- b_m \eta^{\alpha I}
	- 2 V_m{}^{I}{}_J \eta^{\alpha J} \ , \\
\delta_S(\eta) \frak{f}_m{}^a &= \ri \,(\phi_m^I \gamma^a \eta^I) \ .
\end{align}
\end{subequations}

Recall that the vielbein $e_m{}^a$, gravitino $\psi_m{}^{\alpha I}$, dilatation connection
$b_m$ and $\rm SO(\cN)$ connection $V_m{}^{IJ}$ are the fundamental fields, while the
remaining one-forms are composites. The rest of the Weyl multiplet begins to appear
through the transformation law \eqref{eq:dQphiN} of the $\rm SO(\cN)$ connection. There
we see explicitly the bosonic field $w^{IJKL}$ (for $\cN\geq 4$) and the spinor
$w_\alpha{}^{IJK}$ (for $\cN\geq 3$). In the remainder of this appendix, we deal explicitly
with the cases $3 \leq \cN \leq 5$.

{\bf The $\cN$ = 3 case:} \\
As we have discussed, there is a single additional spinor auxiliary field for
$\cN=3$, which can be written $w_\alpha{}^{IJK} = \eps^{IJK} w_\alpha$.
Its supersymmetry and $S$-transformations are
\be
\d_Q (\xi) w_\a = - \frac{1}{2} \eps^{JKL} \xi^\b_J w_{\a\b KL}
	= - \frac{1}{4} \eps^{JKL} (\gamma_c \xi_J)_\a \RN^{c}{}_{KL}|\ , \quad \d_S(\eta) w_\a = 0 \ .
\ee

{\bf The $\cN$ = 4 case:} \\
The auxiliary field sector now involves the scalar fields $w$ and $y$
as well as the spinor field $w_{\alpha I}$. These can be related to the
general definitions \eqref{defCompW} of the fields $w^{IJKL}$, $y^{IJKL}$
and $w_\alpha{}^{IJK}$ via the relations \eqref{eq:N4auxredef}. For
reference, we record their definitions here:
\bsubeq
\begin{align} w &:= \frac{1}{4!} \eps_{IJKL} w^{IJKL} = W| \ , \qquad
	y := \frac{1}{4!} \eps_{IJKL} y^{IJKL} = -\frac{\ri}{4} \nabla^\alpha_I \nabla_\alpha^I W |\ , \\
w_{\a L} &:= \frac{1}{3!} \eps_{IJKL} w_\a{}^{IJK} = - \frac{\ri}{2} \nabla_{\a L} W|~.
\end{align}
\esubeq
Their supersymmetry and $S$-transformations are
\bsubeq
\begin{align}
\delta_Q(\xi) w &= 2 \ri \xi^\alpha_I w_\alpha{}^I~, \qquad \delta_S(\eta) w = 0 \ , \\
\delta_Q(\xi) w_\alpha{}^I &= -\frac{1}{4} \xi_\alpha^I y
	- \frac{1}{2} (\gamma^b \xi^I)_\alpha \hat{\nabla}_b w
	+ \frac{1}{4} \eps^{IJKL} (\gamma^b \xi_J)_\alpha \RN_b{}_{KL}| \ , \\
\delta_S(\eta) w_\alpha{}^I &= \ri \eta_\alpha^I w \ , \\
\delta_Q(\xi) y &= 4 \ri (\xi_I \gamma^a)^\alpha \hat{\nabla}_a w_\alpha{}^I\ , \\
\delta_S(\eta) y &= -4 \eta^{\beta}_J w_\beta{}^J \ ,
\end{align}
\esubeq
where the supercovariant derivatives appearing above are given by
\bsubeq
\begin{align} \hat{\nabla}_a w &:= \cD_a w - \ri \psi_a{}^\a_I w_\a^I \ , \\
\hat{\nabla}_a w_\a{}^I &:= \cD_a w_\a{}^I + \frac{1}{8} \psi_a{}_\a^I y
	+ \frac{1}{4} (\g^b \psi_a{}^I)_\a \hat{\nabla}_b w 
	\non\\ &\qquad
	- \frac{1}{8} (\g^b \psi_a{}_J)_\a \RN_{bKL}| \eps^{IJKL}
	- \frac{\ri}{2} \phi_a{}_\a^I w \ .
\end{align}
\esubeq

{\bf The $\cN$ = 5 case:} \\
The auxiliary field sector is given by
{\allowdisplaybreaks 
\bsubeq
\begin{align} w_I &:= \frac{1}{4!} \eps_{IJKLP} w^{JKLP} = W_I | \ , \\
y_I &:= \frac{1}{4!} \eps_{IJKLP} y^{JKLP} = - \frac{\ri}{3} \nabla^\g_P \nabla_\g^P W_I | \ , \\
w_\a{}^{IJ} &:= \frac{1}{3!} \eps^{IJKLP} w_{\a KLP} = - \frac{\ri}{2} \nabla_\a^{[I} W^{J]}| \ , \\
X_\a &:= \frac{1}{5!} \eps_{IJKLP} X_\a{}^{IJKLP} = \frac{\ri}{5} \nabla_\a^I W_I |~.
\end{align}
\esubeq }
Their transformation rules under supersymmetry and $S$-supersymmetry are
\bsubeq
\begin{align}
\delta_Q(\xi) w^I &= - \ri \xi^{\a I} X_\a + 2 \ri \xi^\alpha_J w_\alpha{}^{JI}~, \qquad \delta_S(\eta) w^I = 0 \ , \\
\d_Q(\xi) w_\a{}^{IJ} &= - \frac{1}{4} \eps^{IJKLP} (\g^b \xi_K)_\a \RN_{b LP}|
- \frac{1}{2} \xi^{[I}_\a y^{J]} - (\g^b \xi^{[I})_\a \hat{\nabla}_b w^{J]} \ , \\
\d_S(\eta) w_\a{}^{IJ} &= 2 \ri \eta_\a^{[I} w^{J]} \ , \\
\d_Q(\xi) X_\a &= - \frac{1}{2} \xi_{\a I} y^I + (\g^b \xi_I)_\a \hat{\nabla}_b w^I~, \qquad \delta_S(\eta) X_\a = - 2 \ri \eta_{\a I} w^I \ , \\
\d_Q(\xi) y^I &= 2 \ri (\g^a)_\a{}^\b \xi^{\a I} \hat{\nabla}_a X_\b + 4 \ri (\g^a)_\a{}^\b \xi^\a_J \hat{\nabla}_a w_\b{}^{JI} - 2 \ri \eps^{IJKLP} \xi^\b_J w_K w_{\b}{}_{LP} \ , \\
\d_S(\eta) y^I &= - 2 \eta^{\b I} X_\b - 4 \eta^\b_J w_\b{}^{JI} \ , 
\end{align}
\esubeq
where we have used the supercovariant derivatives
\bsubeq
\begin{align} \hat{\nabla}_a w^I &:= \cD_a w^I + \frac{\ri}{2} \psi_a{}^{\a I} X_\a - \ri \psi_a{}^\a_J w_\a{}^{JI} \ , \\
\hat{\nabla}_a w_\a{}^{IJ} &:= \cD_a w_\a{}^{IJ} + \frac{1}{8} \eps^{IJKLP} (\g^b \psi_{a K})_\a \RN_{b LP}| + \frac{1}{4} \psi_a{}_\a^{[I} y^{J]} \non\\
&\qquad + \hf (\g^b \psi_a{}^{[I})_\a \hat{\nabla}_b w^{J]}- \ri \phi_a{}^{[I}_\a w^{J]} \ , \\
\hat{\nabla}_a X_\a &:= \cD_a X_\a + \frac{1}{4} \psi_{a \a I} y^I - \frac{1}{2} (\g^b \psi_{a J})_\a \hat{\nabla}_b w^J + \ri \phi_a{}_{\a I} w^I \ .
\end{align}
\esubeq

We have chosen to normalize the component fields at different values of $\cN$
so that the truncation from $\cN=5$ to $\cN=4$, eq.  \eqref{eq:5to4Trunc},
and from $\cN=4$ to $\cN=3$, eq. \eqref{eq:4to3Trunc}, is completely straightforward.

%%%%%%%%%%%%%%%%%%%%%%%%%%%%%%%%%%%%%%%%%%%%%%%%%%%%%%
%%%%%%%%%%%%%%%%%%%%%%%%%%%%%%%%%%%%%%%%%%%%%%%%%%%%%%

\begin{footnotesize}

\end{footnotesize}

\end{document}